\documentclass[fleqn,usenatbib]{mnras}

\usepackage{newtxtext,newtxmath}

\usepackage[T1]{fontenc}

\DeclareRobustCommand{\VAN}[3]{#2}
\let\VANthebibliography\thebibliography
\def\thebibliography{\DeclareRobustCommand{\VAN}[3]{##3}\VANthebibliography}

\usepackage{graphicx}	
\usepackage{amsmath}	
\usepackage{caption}
\usepackage{subcaption}
\usepackage[inkscapearea=page]{svg}

\title[Tidally Perturbed Rotating Stellar Systems]{Tidally Perturbed, Rotating Stellar Systems: Asynchronous Equilibria}

\author[L. A. Z. Arditi and A. L. Varri]{
Lucy A. Z. Arditi$^{1}$\thanks{E-mail: lucy.arditi@ed.ac.uk} and Anna Lisa Varri$^{1,2}$
\\
$^{1}$Institute for Astronomy, University of Edinburgh, Royal Observatory, Blackford Hill, Edinburgh EH9 3HJ, UK\\
$^{2}$School of Mathematics and Maxwell Institute for Mathematical Sciences, University of Edinburgh, King’s Buildings, Edinburgh EH9 3FD, UK
}

\date{Accepted XXX. Received YYY; in original form ZZZ}

\pubyear{2022}

\begin{document}
\label{firstpage}
\pagerange{\pageref{firstpage}--\pageref{lastpage}}
\maketitle

\begin{abstract}
We present a new three-parameter family of self-consistent equilibrium models for quasi-relaxed stellar systems that are subject to the combined action of external tides and rigid internal rotation. These models provide an idealised description of globular clusters that rotate asynchronously with respect to their orbital motion around a host galaxy. Model construction proceeds by extension of the truncated King models, using a newly defined asynchronicity parameter to couple the tidal and rotational perturbations. The method of matched asymptotic expansion is used to derive a global solution to the free boundary problem posed by the corresponding set of Poisson-Laplace equations. We explore the relevant parameter space and outline the intrinsic properties of the resulting models, both structural and kinematic. Their triaxial configuration, characterised by extension in the direction of the galactic centre and flattening toward the orbital plane, is found to depart further from spherical symmetry for larger values of the asynchronicity parameter. We hope that these simplified analytical models serve as useful tools for investigating the interplay of tidal and rotational effects, providing an equilibrium description that complements, and may serve as a basis for, more realistic numerical simulations.
\end{abstract}




\section{Introduction} \label{sec:introduction}
Over the last few decades, we have witnessed a transformation in the quantity and quality of kinematic data available for the characterisation of Galactic globular clusters. This is in large part due to the Gaia astrometric mission \citep{Gaia}, supplemented by high-precision proper motions from the Hubble Space Telescope \citep[see, e.g.,][]{HST2,HST} and radial velocities from spectroscopic instruments such as MUSE, FLAMES, and KMOS \citep[see, e.g.,][] {MUSE,Ferraro}. Most recently, the Euclid Early Release Observation programme has offered deep, wide-field imaging of two Galactic globular clusters \citep{2025A&A...697A...8M}. This has enabled their high-quality morphological study, which is an essential complement to any kinematic investigation. Taken all together, this abundance of detailed phase-space information has revealed globular clusters, once considered the prototypical simple stellar system, to contain significant internal dynamical complexity.

From a morphological perspective, the structural flattening of observed globular clusters was already established decades ago \citep[see, e.g.,][]{Geyer,WhiteShawl}, but renewed systematic efforts are required \citep[see, e.g.,][]{Bergh}. Specifically, we are now able to probe deeper into cluster peripheries, revealing both deviations from spherical symmetry \citep[see, e.g.,][]{Chen} and a wealth of extra-tidal features \citep[see, e.g.,][]{Jordi,Zhang}. Meanwhile, proper motion studies have uncovered rich internal kinematics, including velocity anisotropy and, of particular significance to this work, internal rotation \citep[see, e.g.,][]{Bellini,Bianchini,Sollima,Vasiliev}. Numerous analyses have substantiated the widespread existence of multiple stellar populations within individual globular clusters, distinguished by differences in their light element abundances \citep[for a recent review, see][]{MultiplePops}. Moreover, recent studies have uncovered corresponding kinematic signatures in some clusters, with different stellar populations exhibiting distinct rotational speeds \citep[see, e.g.,][]{Cordero,Dalessandro} and anisotropies in their velocity distributions \citep[see, e.g.,][]{Cordoni,Leitinger}.

The rapid observational progress outlined above motivates a concerted effort on the side of theory to explain this emerging phase space complexity. One approach given renewed impetus is that of equilibrium dynamical modelling \citep[for an introduction, see][]{HeggieHut}. Historically, the description furnished by simple spherical, isotropic, non-rotating models was seen as sufficient to capture salient cluster properties. In particular, the family of \cite{King} models has provided a remarkably good fit to observations \citep[see, e.g.,][]{McLaughlin}. These models are defined by an ergodic\footnote{In this context, ergodic refers to a distribution function that is a function of energy only, and therefore, by Jeans’ theorem, must be a steady-state solution to the collisionless Boltzmann equation.} one-body distribution function that is defined by introducing a truncation, continuous in phase space, to the isothermal sphere. The truncation acts as a zeroth-order approximation to the gravitational field of a host galaxy, neglecting all tidal and inertial forces, and results in a lowered Maxwellian distribution function that describes a stellar system in a quasi-relaxed state. This is physically justified for Galactic globular clusters, whose ages typically exceed their two-body relaxation times. Self-consistency is achieved, under the assumption that the stellar component constitutes the total mass of the system, by solving the Poisson equation for the mean-field cluster potential that appears in the definition of the distribution function.

The \cite{King} models, despite their erstwhile success, cannot capture the morphological and kinematic complexity that is now widely observed. However, the idealisation inherent to equilibrium modelling allows for the introduction and controlled analysis of additional physical ingredients. Useful insight for the interpretation of recent observational results may therefore be obtained from more physically detailed models, including those devised as direct extensions to the \cite{King} models.

The \cite{King} models, as well as other ergodic distribution functions, may be modified to include the action of an external tidal field \citep{Weinberg,HeggieRamamani,VarriBertin1} via the appropriate generalised energy invariant (Jacobi integral; for details, see Section~\ref{sec:construction}). Explicitly including tidal effects beyond the spherical truncation breaks the symmetry of the original equilibria. By this method, \cite{VarriBertin1} generate a family of self-consistent, non-spherical models to describe a quasi-relaxed stellar system on a circular orbit within a host galaxy (represented by a static, spherically-symmetric potential). The model properties are discussed in \cite{VarriBertin2}. An approximate solution to the Poisson equation is found by the method of matched asymptotic expansion \citep[see][]{Smith}, a technique frequently adopted to study the structure of rotating \citep{Chandra1} and tidally-perturbed \citep{Chandra2,Chandra3} polytropic stars. \cite{HeggieRamamani} use a different perturbative method of solution to treat the same distribution function, but their analysis concerns only maximally perturbed models. In \cite{VarriBertin3}, the general procedure developed in \cite{VarriBertin1} is applied to the axisymmetric problem of rigid internal rotation.

While these models enable the structural and kinematic effects of external tides and internal rotation to be separately examined, there currently exists no equilibrium model capable of treating the interplay of these two physical ingredients. The tidal problem set out in \cite{VarriBertin1} implicitly assumes that the cluster rotates synchronously with its orbital motion, i.e., that the corresponding angular velocities are equal. The more general case would allow the rotational and orbital motion to be decoupled, resulting in asynchronous cluster rotation inside a tidal field. The case for this generalised scenario is threefold: first, as discussed, rotation is now measured in several Galactic globular clusters (e.g., see \citealt{Bianchini,Vasiliev} for recent, comprehensive investigations); second, analytic and numerical studies \citep[see, e.g.,][]{Keenan,FukushigeHeggie,Ernst} have found that stars on prograde orbits are preferentially lost, suggesting evaporative mass loss is accompanied by the emergence of net retrograde rotation; and finally, there is evidence from N-body simulations that both initially non-rotating and synchronous clusters evolve in a tidal field into a state of partial, not full, synchronisation \citep[see][and subsequent studies]{Tiongco}.

A mathematically equivalent problem is found in the theory of binary stars, when one component deviates from synchronism between its axial rotation and orbital motion. This physical model was first considered in \cite{Kopal}, with subsequent treatments by \cite{Plavec}, \cite{Kruszewski}, \cite{Limber}, \cite{Naylor}, \cite{Lubow}, and \cite{Sepinsky}. \cite{AvniSchiller} further extends this analysis to describe a misaligned binary system, where the rotation axis of one component is not parallel to the axis of orbital revolution. Since these studies were concerned with mass transfer in close binaries, they focus on characterising the external potential with a simple Roche model \citep[for an introduction, see][]{Horedt}, and neglect the internal stellar structure. \cite{Naylor} makes a comparison between this approach and a polytropic model \citep{NaylorAnand} adapted from \cite{Chandra1}. This required a numerical factor defined in \cite{Kopal} to quantify the deviation between rotational and orbital angular frequencies to be treated as a free parameter, thus simplifying the perturbative analysis involved.

Here, we employ the method of matched asymptotic expansion to generate a new family of self-consistent equilibrium models that describe a quasi-relaxed stellar system with rigid internal rotation that is asynchronous with respect to its orbital motion inside a tidal field. This article is organised as follows. In Section~\ref{sec:construction}, we introduce the physical model and derive a distribution function by extension of the \cite{King} models. We introduce the Poisson and Laplace equations that determine the potential internal and external to the modelled system. Relevant parameters are defined in Section~\ref{sec:parameter} and the resulting parameter space is explored. In Section~\ref{sec:asymptotic_matching}, a global solution to the system of Poisson-Laplace equations is obtained by asymptotic matching. Details of the analytic calculations and numerical implementation are presented in Appendices~\ref{sec:A} and \ref{sec:BTIRO}. Intrinsic structural and kinematic properties of the models thus constructed are presented in Section~\ref{sec:properties}. We conclude in Section~\ref{sec:conclusions} with a summary and discussion of the results obtained.

\section{Model Construction} \label{sec:construction}

The idealised physical model consists of a globular cluster characterised by rigid internal rotation that is asynchronous with respect to its orbital motion around a host galaxy. The cluster's centre of mass is taken to follow a circular orbit of galactocentric radius $R=R_\text{0}$. The gravitational field of the host galaxy is described by a static, spherically-symmetric potential $\Phi_{\text{G}}=\Phi_{\text{G}}(R)$. The orbital and rotational angular frequencies, $\Omega$ and $\omega$, are, in general, not equal. In the following calculations, attention is confined to the case where the rotational and orbital axes are aligned.

\subsection{Perturbation potential}
\label{sec:potential}

In the case of synchronous rotation, \cite{VarriBertin1} define a local frame with origin at the cluster's centre of mass. This frame, described by Cartesian coordinates $(x,y,z)$, rotates with the cluster such that the $x$-axis always points away from the galactic centre, the $y$-axis points in the direction of orbital motion, and the $z$-axis is perpendicular to the orbital plane. Relative to this frame, an asynchronously rotating cluster will rotate with non-zero angular frequency $\omega-\Omega$ about the $z$-axis.

We define a new local frame that coincides with this synchronous frame at the origin and along the rotation axis, but which rotates with the cluster such that the bulk rotation vanishes. The position vector in the corotating frame is given by $\mathbfit{r}=(\alpha,\beta,z)$, where $\alpha = x\cos\eta+y\sin\eta $ and $ \beta = y\cos\eta -x\sin\eta$. Here $\eta=(\omega-\Omega)t$ is the angle between the $\alpha$- and $x$- (or, equivalently, $\beta$- and $y$-) axes at time $t$, specified up to an arbitrary constant. The specific Lagrangian for a star belonging to the cluster is
\begin{align}
    \mathcal{L}=&\frac{1}{2}\bigg\{\dot{\alpha}^2+\dot{\beta}^2+\dot{z}^2 +2\omega\left(\alpha\dot{\beta}-\dot{\alpha}\beta\right) + 2\Omega R_\text{0}\dot{y} \notag  + \left(\omega^2-\Omega^2\right)\\&\times\left(\alpha^2+\beta^2\right) +\Omega^2\left[\left(R_\text{0}+x\right)^2+y^2\right]\bigg\} - \Phi_{\text{G}}(R) - \Phi_{\text{C}}(\mathbfit{r},t), \label{eq:lagrangian}
\end{align}
where $\Phi_{\text{C}}(\mathbfit{r},t)$ is the mean-field potential of the cluster. This is a periodic function of time in the corotating frame, of period $T\!=\!2\pi/(\omega\!-\!\Omega)$, due to the variation of the galactic potential as the cluster rotates. Note that the dependence of $x$ and $y$ on angle $\eta$ mean they too are periodic functions of time.

The equations of motion for an individual star can be obtained either directly from the Lagrangian or by transforming into corotating coordinates those obtained in the synchronous case \citep[see, e.g.,][equations 2--4 therein]{VarriBertin1}. These are the equations of motion for orbits in the Hill's approximation, a limiting case of the restricted, circular three-body problem where the primary body is significantly more massive than the secondary, and the tertiary's mass is negligible compared to the other two \citep{Hill}. The spatial derivatives of $\Phi_{\text{G}}$ are linearised under the above approximation, by assuming the dimensions of the cluster to be much smaller than the orbital radius ($\alpha,\beta,z\ll R_\text{0}$). The equations of motion of a star in the corotating frame are
\begin{equation}
    \ddot{\alpha}-2\dot{\beta}\omega+\alpha\left(\Omega^2-\omega^2\right) -\Omega^2\nu x\cos\eta = -\frac{
    \partial\Phi_{\text{C}}}{\partial \alpha}, \label{eq:eomalpha}
\end{equation}
\begin{equation}
    \ddot{\beta}+2\dot{\alpha}\omega+\beta\left(\Omega^2-\omega^2\right) -\Omega^2\nu x\sin\eta = -\frac{
    \partial\Phi_{\text{C}}}{\partial \beta}, \label{eq:eombeta}
\end{equation}
\begin{equation}
    \ddot{z}+\Omega^2z = -\frac{\partial \Phi_{\text{C}}}{\partial z}. \label{eq:eomz}
\end{equation}
The value of the dimensionless coefficient $\nu=4-\kappa^2/\Omega^2$ is determined by the galactic potential through the orbital frequency $\Omega^2=(d\Phi_G/dR)_{R_\text{0}}/R_\text{0}$ and epicyclic frequency $\kappa^2=3\Omega^2+(d^2\Phi_\text{G}/dR^2)_{R_\text{0}}$.

Due to the time-dependence of the galactic tidal field (and, consequently, cluster potential), equations~(\ref{eq:eomalpha})--(\ref{eq:eomz}) do not admit an exact integral of motion. The quantity
\begin{equation}
H = \frac{1}{2}\left(\dot{\alpha}^2+\dot{\beta}^2+\dot{z}^2\right)+\Phi_\text{P}(\mathbfit{r},t)+\Phi_\text{C}(\mathbfit{r},t), \label{eq:iom}
\end{equation}
is conserved only in the limit of synchronicity, for which the explicit time-dependence vanishes. Here, we define a perturbation potential
\begin{equation}
    \Phi_\text{P}(\mathbfit{r},t)=\frac{1}{2}\left(\Omega^2-\omega^2\right)\left(\alpha^2+\beta^2\right)+\frac{1}{2}\Omega^2\left(z^2-\nu x^2\right), \label{eq:potential}
\end{equation}
or equivalently
\begin{equation}
    \Phi_\text{P}(x,y,z) = \Phi_{\text{T}}(x,z) + \Phi_{\text{cen}}(x,y) + \frac{1}{2}\Omega^2\left(x^2+y^2\right), \label{eq:potential2}
\end{equation}
where $\Phi_\text{T}(x,z)=\Omega^2(z^2-\nu x^2)/2$ and $\Phi_\text{cen}(x,y)=-\omega^2(x^2+y^2)/2$ are the tidal and centrifugal potentials identified in the synchronous case and the case of rigid rotation without a tidal field. The former is due to orbital motion in an external tidal field, while the latter consists of centrifugal terms from internal rotation. The final term in equation~(\ref{eq:potential2}) is similar in form to the centrifugal potential, and it arises from the gradient of $\Phi_\text{G}$ in the corotating frame. Note that all Coriolis terms have cancelled, hence do not appear in the adopted form of the perturbation potential.

In the general case, we have
\begin{equation}
    \frac{dH}{dt}=\frac{\partial(\Phi_P(\mathbfit{r},t)+\Phi_C(\mathbfit{r},t))}{\partial t}. \label{eq:dHdt}
\end{equation}
However, we may assume that dynamical timescales within the cluster are much shorter than the timescale of axial rotation in the synchronous local frame. Under this assumption, equation~(\ref{eq:iom}) may be treated as a conserved quantity over dynamical timescales, for which the right hand side of equation~(\ref{eq:dHdt}) approaches zero. Proceeding in this way, we associate equation~(\ref{eq:iom}) with the Jacobi integral defined in the classical formulation of the restricted three-body problem. This quantity comprises the energy and angular momentum as defined in an inertial frame centred on the cluster's barycentre, which are not individually conserved, and can be physically interpreted as the specific single-star energy in the corotating frame.

The time-dependence of the perturbation potential is an essential difference between the models presented here and those that describe stellar systems in synchronous rotation \citep[see, e.g.,][]{HeggieRamamani,VarriBertin1}. This periodic variation is formally inconsistent with the premise of rigid rotation, inducing bulk radial oscillations in the modelled cluster. These motions are negligible over the dynamical timescales we have adopted  \citep[see also the discussions in][]{Limber,Todoran,Sepinsky}; therefore, we proceed with the assumption of uniform circular rotation, noting this simplification in the resulting models.

\subsection{Distribution function} \label{sec:df}

The lowered Maxwellian distribution function that defines the \cite{King} models, $f_\text{K}(E)$, is extended to describe the case of asynchronous rotation in a tidal field. This is done by replacing $E$, the single-star energy for an isolated, non-rotating cluster, with $H$, the single-star energy in the co-rotating frame given in equation~(\ref{eq:iom}),
\begin{equation}
    f_\text{K}(H)=
\begin{cases} 
      A\left[e^{-aH}-e^{-aH_{0}}\right] & H\leq H_{0}\\
      0 & H > H_{0}.\\ 
   \end{cases} \label{eq:df}
\end{equation}
$A$ and $a$ are positive scaling constants. The energy spectrum is truncated at the cut-off value $H_0$, above which a star becomes energetically unbound. This truncation results in a spatially finite model, bounded by the equipotential surface, in the total effective potential $\Phi_\text{P}+\Phi_\text{C}$, at which the escape velocity vanishes. Defining the dimensionless escape energy as
\begin{equation}
    \psi(\mathbfit{r},t)=a\left[H_{0}-\Phi_\text{P}(\mathbfit{r},t)-\Phi_\text{C}(\mathbfit{r},t)\right], \label{eq:escape}
\end{equation}
such a boundary surface is implicitly defined by $\psi(\mathbfit{r},t)=0$. Within the cluster, $\psi$ takes positive values up to the central value $\Psi=\psi(\mathbfit{0},t)$.

The shape of the cluster boundary manifestly depends on both the cluster potential $\Phi_\text{C}$ and the perturbation $\Phi_\text{P}$. While the former is unknown a priori, the structure of the latter suggests a triaxial geometry. For synchronous models, the tidal perturbation produces (for positive $\nu$) compression in the $z$-direction and extension in the $x$-direction \citep{HeggieRamamani,VarriBertin1}. The additional axisymmetric terms in $\Phi_\text{P}$ do not break the symmetry of this result, but rather introduce an additional uniform extension or compression in the equatorial plane. For $|\omega|<|\Omega|$, the perturbation potential is raised relative to the synchronous case ($\omega=\Omega$), suggesting compression, while extension is expected for $|\omega|>|\Omega|$. 

The number and location of saddle points in the total potential depend similarly on the relative magnitudes of the rotational and orbital frequencies. For $|\omega|\leq|\Omega|$, the system has two saddle points, positioned on the $x$-axis symmetrically about the origin. This pair of saddle points is analogous to the two Lagrange points identified in Hill's problem. For $|\omega|>|\Omega|$, as the system approaches the limit of pure internal rotation ($\Omega\!\rightarrow\!0$), the saddle points form an infinite set distributed as a continuous circle in the equatorial plane. In either case, the distance from the origin to the saddle point located on the positive $x$-axis is called the tidal, or Jacobi, radius $r_\text{T}$ and satisfies
\begin{equation}
    \frac{\partial}{\partial x} \psi(x=r_\text{T},y=0,z=0)=0. \label{eq:rT}
\end{equation}

Since the structure of the distribution function of the \cite{King} model is retained, the usual analytic expression for the density is obtained by integrating over the velocity space \citep[see, e.g.,][]{HeggieHut},
\begin{equation}
    \rho (\psi) = \hat{A}e^\psi\gamma{\left(\frac{5}{2},\psi\right)}=\hat{A}\hat{\rho}(\psi). \label{eq:density}
\end{equation}
where $\gamma$ is the lower incomplete gamma function, $\hat{\rho}(\psi)$ denotes the dimensionless density, and $\hat{A}=(8\sqrt{2}\pi A/3a^{3/2})e^{-aH_0}$ is a normalising constant fully specified by the values of $A$, $a$ and $H_0$. The velocity dispersion, which is isotropic for all ergodic distribution functions, is given by the second moment of the distribution function in velocity space,
\begin{equation}
    \sigma^{2} (\psi) = \frac{2}{5a}\frac{\gamma \left(7/2,\psi\right)}{\gamma \left(5/2,\psi\right)} = \frac{1}{a}\hat{\sigma}^2(\psi), \label{eq:dispersion}
\end{equation}
where $\hat{\sigma}$ denotes the dimensionless velocity dispersion.

\subsection{Poisson-Laplace equations}
\label{sec:poisson}
To fully construct a self-consistent family of models defined by equation~(\ref{eq:df}), we must solve the corresponding Poisson equation for the internal cluster potential. In terms of the dimensionless escape energy, we have
\begin{equation}
   \hat{\nabla}^2\left[\psi+a\Phi_\text{T}+a\Phi_\text{cen}+ a\frac{1}{2}\Omega^2r_0^2\left(\hat{x}^2+\hat{y}^2\right)\right]=-9\frac{\hat{\rho}(\psi)}{\hat{\rho}(\Psi)}. \label{eq:p1}
\end{equation}
From here onwards, we adopt a dimensionless coordinate system by rescaling all spatial variables with respect to the King radius $r_{0}=(9/4\pi aG\rho_0)^{1/2}$, a scale length that depends on the central density $\rho_0=\rho(\Psi)$. The position vector takes dimensionless form $\hat{\mathbfit{r}}\!=\!(\hat{\alpha},\hat{\beta},\hat{z})\!=\!\mathbfit{r}/r_0$, with magnitude $\hat{r}=|\hat{\mathbfit{r}}|$, and the dimensionless Laplacian operator is given by $\hat{\nabla}^2 =r_0^2\nabla^2$. For the models that separately treat pure internal rotation and synchronous rotation in a tidal field, the centrifugal and tidal potentials are written in dimensionless form as $a\Phi_\text{cen}\equiv\chi C$ and $a\Phi_\text{T}\equiv\epsilon T$. Here $\chi$ and $\epsilon$ are the dimensionless rotational and tidal strength parameters,
\begin{equation}
    \chi=\frac{\omega^2}{4\pi G\rho_0}, \label{eq:chi}
\end{equation}
\begin{equation}
    \epsilon=\frac{\Omega^2}{4\pi G\rho_0}. \label{eq:epsilon}
\end{equation}
These parameters allow us to define the corresponding dimensionless quadratic functions $C= -(9/2)(\hat{x}^2+\hat{y}^2)$ and $T= (9/2)(\hat{z}^2-\nu\hat{x}^2)$. The remaining perturbation term in equation~(\ref{eq:p1}), which has the same form as $a\Phi_\text{cen}$ but with $-\Omega$ in place of $\omega$, may be written as $-\epsilon C$. Therefore, after taking the Laplacian of these terms and rearranging, the dimensionless Poisson equation~(\ref{eq:p1}) becomes
\begin{equation}
     \hat{\nabla}^2\psi= -9\left[\frac{\hat{\rho}(\psi)}{\hat{\rho}(\Psi)}+(1-\nu)\epsilon-2(\chi-\epsilon)\right]. \label{eq:poisson}
\end{equation}

Equation~(\ref{eq:poisson}) governs the cluster potential in the internal domain (i.e., within the boundary of the cluster), where the density $\hat{\rho}$ does not vanish. The corresponding prescription for the external domain is given by the dimensionless Laplace equation, which lacks the source term,
\begin{equation}
    \hat{\nabla}^2\psi = -9\left[(1-\nu)\epsilon-2(\chi-\epsilon)\right]. \label{eq:laplace}
\end{equation}
Since the boundary surface, which is defined by $\psi=0$, can not be found without first knowing the structure of the cluster potential, finding a global solution to this pair of elliptic partial differential equations constitutes a free boundary problem. The relevant set of boundary conditions is as follows: at the origin, the solution must satisfy\footnote{Although $\psi$ is explicitly time-dependent in the co-rotating frame, we note that, by definition of the reference frame itself, the boundary conditions at the origin are constant in time.}
\begin{equation}
    \psi(\mathbfit{0},t)=\Psi, \label{eq:bc1}
\end{equation}
and
\begin{equation}
    \hat{\nabla}\psi(\mathbfit{0},t) = \mathbfit{0}, \label{eq:bc2}
\end{equation}
and at large radii we require $a\Phi_\text{C}\rightarrow0$, or, equivalently,
\begin{equation}
    \psi+\epsilon T +(\chi-\epsilon)C \rightarrow aH_0. \label{eq:bc3}
\end{equation}

\section{Parameter space} \label{sec:parameter}

\subsection{Key definitions} \label{sec:parameter_def}
The family of models thus constructed are characterised by two scale factors, $A$ and $a$, and three dimensionless parameters, $(\Psi,\chi,\epsilon)$. The dimensionless central depth of the potential well, $\Psi$, provides a measure of cluster concentration. This concentration parameter is fixed by the (freely chosen) energy cut-off $H_0$. The rotational strength parameter, $\chi$, defined in equation~(\ref{eq:chi}), corresponds to the squared ratio of the rotational angular frequency to the dynamical angular frequency associated with the central density $\rho_0$. The strength of the tidal perturbation is quantified similarly by the tidal strength parameter $\epsilon$, given by equation~(\ref{eq:epsilon}). Synchronous rotation is obtained for $\chi=\epsilon$, while an isolated rotating model is recovered in the limit of vanishing tidal perturbation ($\epsilon\rightarrow0$), i.e. rotational dominance. Although it is tempting to construct a single perturbation parameter $K=2\chi+\epsilon(\nu-3)$ from the terms that appear on the right-hand side of equation (\ref{eq:poisson}), this would contain insufficient information to fully characterise the model. As seen in Section~\ref{sec:poisson}, the contribution to the total perturbation from $T(\hat{x},\hat{z})$ depends on $\epsilon$, whereas the contribution from $C(\hat{x},\hat{y})$ depends on the difference $\chi-\epsilon$. Therefore, two separate parameters are required. 

The problem can, however, be reduced to a single perturbation expansion if we disregard models in the regime of rotational dominance. This may be justified by reference to observed globular clusters, which do not necessarily rotate but are invariably embedded within a host galaxy. This enables us to adopt the notation first used in \citet[][]{Kopal}, whereby the orbital and rotational angular frequencies are related by a coefficient $f$,
\begin{equation}
    \omega = (1+f)\Omega. \label{eq:f}
\end{equation}
While the models can no longer be reduced to the purely rotational limit (since $\omega=0$ if $\Omega=0$), synchronous rotation is obtained by taking $f\!=\!0$. Relative to an inertial frame centred on the cluster's barycentre, $f\!=\!-1$ corresponds to the absence of internal rotation ($\omega=0$), while $f\!<\!-1$ and $f\!>\!-1$ give retrograde ($\omega<0$) and prograde\footnote{We define prograde (retrograde) rotation as rotation in the same (opposite) sense, with respect to the inertial frame, as the cluster's orbital motion around the host galaxy.} ($\omega>0$) rotation, respectively. Using equation~(\ref{eq:f}), we can further define define the asynchronicity parameter $\zeta= f^2+2f$, such that
\begin{equation}
    \chi-\epsilon=\zeta\epsilon. \label{eq:zeta}
\end{equation}
Therefore, the Poisson equation (\ref{eq:poisson}) becomes
\begin{equation}
     \hat{\nabla}^2\psi= -9\left[\frac{\hat{\rho}(\psi)}{\hat{\rho}(\Psi)}+(1-\nu-2\zeta)\epsilon\right], \label{eq:newpoisson}
\end{equation}
with the following boundary condition at large radii
\begin{equation}
    \psi+\epsilon(T+\zeta C) \rightarrow aH_0. \label{eq:combbound}
\end{equation}
By adopting this formulation, we have chosen to introduce the rotational perturbation at first order in $\epsilon$. If we now treat $\zeta$ as a free parameter, we may construct the models using a single perturbation expansion in $\epsilon$.

We note that $\zeta$, as a function of  $f$, is symmetric about the minimum\footnote{Although $\zeta$ has no formal maximum value, we limit our investigation to small deviations from synchronicity. As discussed in Section~\ref{sec:potential}, we have neglected ordered, periodic motions resulting from the time variation of the cluster potential. We expect the excess radial velocity associated with these oscillations to be maximised for critical models (see Section \ref{sec:regimes}) and to increase in magnitude with the absolute value of $\zeta$. The assumption of uniform rotation, therefore, becomes increasingly inaccurate as the value of $\zeta$ increases.} $\zeta(-1)\!=\!-1$, which corresponds to the absence of rotation in the inertial frame. The structure of asynchronous models must therefore be independent of whether rotation is prograde or retrograde, with the asynchronicity parameter depending only on the relative magnitudes of the angular frequencies: $\zeta>0$ for $|\omega|>|\Omega|$, $\zeta<0$ for $|\omega|<|\Omega|$ and $\zeta=0$ for $|\omega|=|\Omega|$. This last condition corresponds to either synchronous rotation, where $\omega=\Omega$, or what might be termed antisynchronous rotation, where $\omega=-\Omega$. This was already apparent from the discussion in Section~\ref{sec:df}. In conclusion, we are free to use $\zeta$ (in place of $\chi$) to characterise the internal rotation of the models, with the three main degrees of freedom of the models now set by the parameters $(\Psi,\zeta,\epsilon)$.

\subsection{Perturbation regimes} \label{sec:regimes}

\begin{figure}
	\includegraphics[width=\columnwidth]{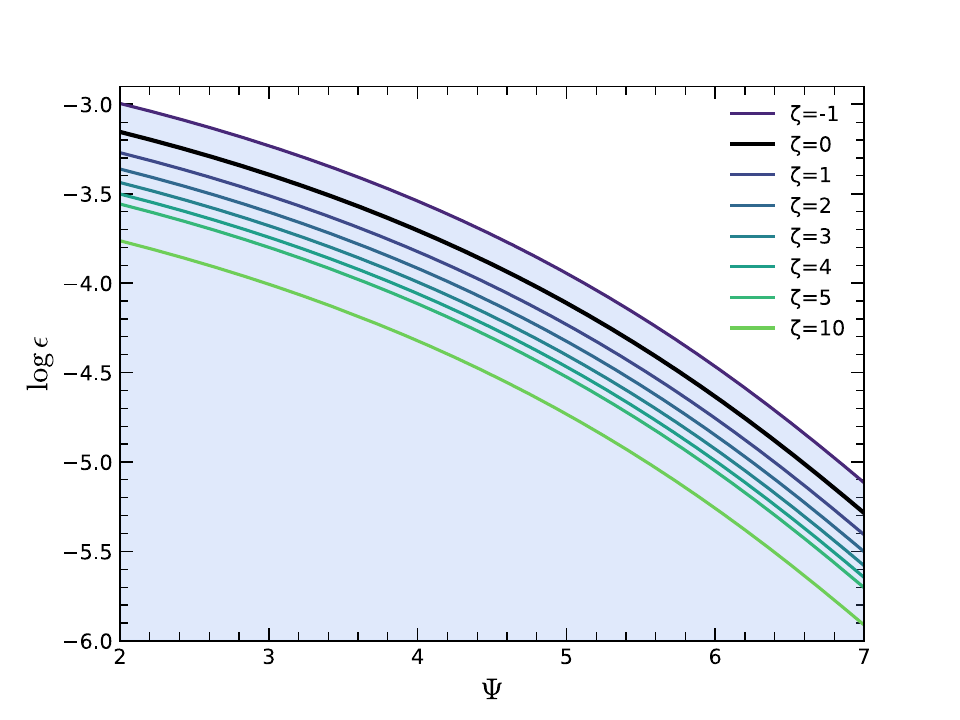}
    \caption{The ($\Psi\!,\!\epsilon$) parameter space for first-order models with $\nu=3$. Solid lines denote the critical values of the tidal strength parameter $\epsilon_\text{cr}$, as a function of the concentration parameter $\Psi\in[2,7]$, computed for representative values of the asynchronicity parameter $\zeta\in\{-1,0,1,2,3,4,5,10\}$. The shaded area illustrates the maximum extent of the allowed parameter space (see Section \ref{sec:regimes} for details).}
    \label{fig:parameter1}
\end{figure}
\begin{figure}\includegraphics[width=\columnwidth]{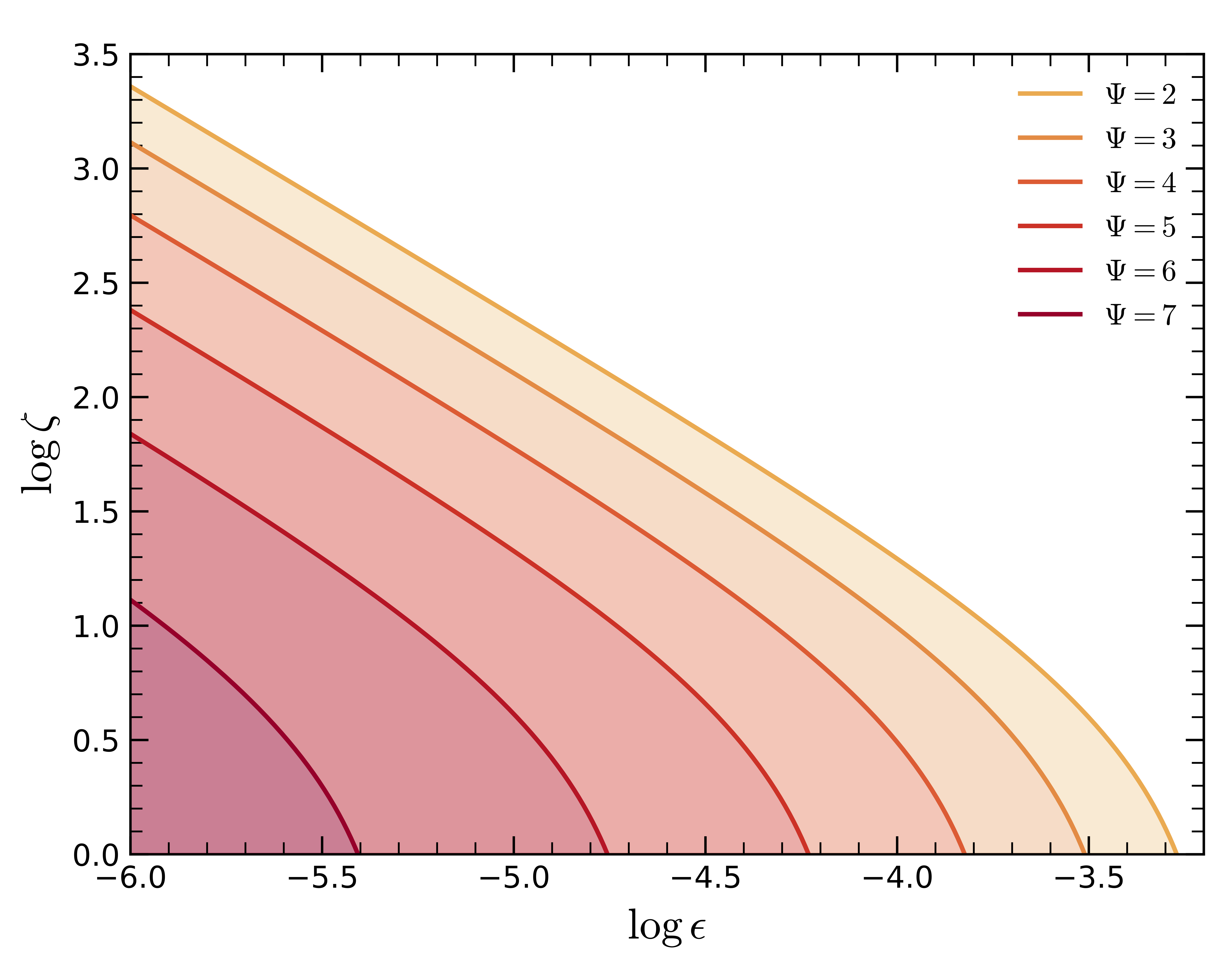}
    \caption{The ($\epsilon\!,\!\zeta$) parameter space for first-order models with $\nu=3$. Solid lines denote the critical values of $(\epsilon,\zeta)$, computed for representative values of the concentration parameter $\Psi\in\{2,3,4,5,6,7\}$. The shaded areas show the parameter space that is accessible at each value of concentration.}
    \label{fig:parameter2}
\end{figure}

The outermost closed equipotential surface for a given model, which contains the saddle points (see equation (\ref{eq:rT}), is known as the critical surface. We call a model critical if its boundary coincides with the critical surface, which is defined implicitly by
\begin{equation}
    \psi(\hat{x}=\hat{r}_\text{T},\hat{y}=0,\hat{z}=0; \epsilon_\text{cr})=0. \label{eq:critical}
\end{equation}
Critical models exhibit the maximum deviation from spherical symmetry possible for a bound stellar system. For given values of the concentration and asynchronicity parameters, the value of the tidal strength parameter that results in a critical model, $\epsilon_\text{cr}$, may be found by numerically solving the system composed of equations~(\ref{eq:rT}) and (\ref{eq:critical}). Larger values of $\epsilon$ result in unbound models, whose boundaries follow open equipotential surfaces. These models are not relevant to our discussion of equilibrium systems.

The $(\Psi,\epsilon)$ parameter space for first-order models with $\nu=3$ (i.e., a Keplerian galactic potential) is illustrated in Fig.~\ref{fig:parameter1}, where the solid lines denotes the critical values $\epsilon_\text{cr}(\Psi)$, computed for different values of the asynchronicity parameter $\zeta$. For a given value of $\zeta$, the allowed parameter space of bound equilibrium models lies below the corresponding solid line. Minimally distorted, sub-critical models are found furthest towards the bottom left. These models are underfilled (i.e., their maximum extension in the equatorial plane is smaller than the tidal radius, $r_T$), and their boundaries follow quasi-spherical equipotential surfaces, which are well embedded inside the critical surface.

The critical values $\epsilon_\text{cr}(\Psi)$ computed for $\zeta=0$ (thick black line) bound the parameter space allowed for synchronous models. Positive values of $\zeta$ are seen to reduce the available parameter space, i.e. they require lower critical values of $\epsilon$ across the explored range of $\Psi$. Conversely, the parameter space is extended for $\zeta<0$, with the critical values of $\epsilon$ for nonrotating models ($\zeta=-1$) defining the maximum possible extent of the accessible parameter space (shaded region). This behaviour may be interpreted in terms of the forces acting against the system's self-gravity: stronger centrifugal forces from increased rotation ($\zeta>0$) lower the threshold at which tidal forces can disrupt the system, while decreased rotation ($\zeta<0$) increases this threshold. Note that, at higher positive values of $\zeta$, the relative magnitude of this effect decreases.

The dependence of $\epsilon_\text{cr}$ on $\zeta$ is illustrated in Fig.~\ref{fig:parameter2}, where we present the $(\epsilon,\zeta)$ parameter space for concentrations between $\Psi=2$ and $\Psi=7$. Sub-critical models are again found towards the bottom left of the plot, with an increase in either the tidal perturbation or asynchronous rotation (keeping other parameters fixed) moving the model towards its critical state. We can see that less concentrated systems can sustain significantly greater rotation strength under the same tidal perturbation. This is consistent with such models having a denser outer envelope, whose stronger self-gravity provides a higher threshold for disruption. 

\section{Solution by Asymptotic Matching} \label{sec:asymptotic_matching}

To obtain a global solution to the Poisson-Laplace equations across the unknown boundary, we employ the method of matched asymptotic expansion. For a given value of $\zeta$, we separately expand the internal (Poisson) and external (Laplace) solutions as asymptotic series in the perturbation parameter $\epsilon$,
\begin{equation}
\psi(\hat{\mathbfit{r}},t;\epsilon)=\sum_{\text{k=0}}^\infty\frac{1}{k!}\psi_k(\hat{\mathbfit{r}},t)\epsilon^k. \label{eq:expansion}
\end{equation}
This series expansion requires $\epsilon$ to be small, and spherical symmetry is assumed for zeroth-order terms (as the unperturbed solution corresponds to the dimensionless escape energy of the spherical \citealt[][]{King} model). We proceed by comparing coefficients between the internal and external series expansions. However, this matching cannot be performed directly at $\psi=0$. Specifically, we note that there is a region of the internal domain (surrounding the boundary surface $\psi=0$) in which the correcting terms $\psi_k$ dominate over the leading term $\psi_0$. We can treat this region as a boundary layer. As the perturbation is singular within this region (i.e., the series is formally disordered), we follow \citet[][]{Smith} in constructing an expansion of the solution in the boundary layer itself using appropriately rescaled coordinates (see Section \ref{sec:boundary} for details). The asymptotic matching is then conducted pairwise between the boundary layer solution and the internal and external solutions. The full solution to first order in $\epsilon$ is presented below, with further details available in Appendix~\ref{sec:A}.

\subsection{Internal region} \label{sec:internal}

Substituting equation~(\ref{eq:expansion}) into the Poisson equation~(\ref{eq:newpoisson}) generates an infinite set of radial Cauchy problems subject to the boundary conditions given in equations~(\ref{eq:bc1})--(\ref{eq:bc2}). The unperturbed ($k=0$) problem is specified by the Poisson equation
\begin{equation}
   \frac{1}{\hat{r}^2}\frac{d}{d\hat{r}}\left(\hat{r}^2\frac{d\psi^\text{(int)}_0}{d\hat{r}}\right) = -9\frac{\hat{\rho}\left(\psi_0^{\text{(int)}}\right)}{\hat{\rho}(\Psi)}, \label{eq:kingpoisson}
\end{equation}
and boundary conditions\footnote{Here and in the following subsections, the prime denotes the derivative with respect to  $\hat{r}$.} $\psi_0^{\text{(int)}}(0)=\Psi$ and $\psi_0^{\text{(int)}\prime}(0)=0$, which, by construction, define the Poisson problem for the spherically symmetric \citet[][]{King} models (see equation 16 therein).

The Poisson equation for the first-order ($k=1$) term constitutes a modification to the synchronous case by the addition of a constant term $18\zeta$ (see Appendix~\ref{sec:Ainternal}). We proceed with an expansion in spherical harmonics\footnote{We use real spherical harmonics with Condon-Shortley phase.},
\begin{equation}
    \psi_1^{\text{(int)}}(\hat{\mathbfit{r}},t)=f_{00}(\hat{r})+\sum_{l=1}^\infty\sum_{m=-l}^l A_{lm}\mathcal{\gamma}_l(\hat{r})Y_{lm}(\theta,\phi). \label{eq:internal}
\end{equation}
The original three-dimensional partial differential equation has been converted into a set of one-dimensional ordinary differential equations in dimensionless radius $\hat{r}$. Note that the time-dependence has been absorbed into the azimuthal angle $\phi$, which is defined with respect to the $\hat{x}$-axis.

The $l=0$ harmonic, $Y_{00}=1/\sqrt{4\pi}$, is incorporated in $f_{00}= Y_{00}\psi_{1,00}^\text{(int)}$, which is governed by the differential equation
\begin{equation}
   \mathcal{D}_0f_{00}=-9(1-\nu-2\zeta). \label{eq:ode1}
\end{equation}
The differential equations for higher degree harmonics ($l\geq1$) are homogeneous,
\begin{equation}    \mathcal{D}_l\psi_{1,lm}^\text{(int)}=0. \label{eq:ode2}
\end{equation}
At all degrees, $\mathcal{D}_l$ is the linear differential operator defined in equation~(\ref{eq:Dl}).
The form of equation~(\ref{eq:ode2}) implies
\begin{equation}
    \psi_{1,lm}^\text{(int)}(\hat{r})=A_{lm}\mathcal{\gamma}_l(\hat{r})
\end{equation}
where $\mathcal{\gamma}_l(\hat{r})\sim\hat{r}^l$ for $\hat{r}\!\rightarrow\!0$ and $A_{lm}$ is an undetermined constant. The boundary conditions for all harmonics ($l\geq0$) are homogeneous,
\begin{equation}
    \psi_{1,lm}^\text{(int)}(0)=\psi_{1,lm}^\text{(int)}\prime(0)=0. \label{eq:internalBCs}
\end{equation}
The $k=1$ problem thus specified is seen to differ from the corresponding synchronous problem only through the constant term in equation~(\ref{eq:ode1}) \citep[see][equation 26]{VarriBertin1}.

\subsection{External region} \label{sec:external}
The general solution to the Laplace equation~(\ref{eq:laplace}) is
\begin{equation}
    \psi^\text{(ext)}(\hat{\mathbfit{r}},t)=\alpha-\frac{\lambda}{\hat{r}}-\sum_{l=1}^{\infty}\sum_{m=-l}^l\frac{\beta_{lm}}{\hat{r}^{l+1}}Y_{lm}(\theta,\phi) -\epsilon\big[T(\hat{\mathbfit{r}},t)+\zeta C(\hat{\mathbfit{r}},t)\big].
\end{equation}
This is composed of the particular solution $-\epsilon(T+\zeta C)$ and 
the general solution to the homogeneous equation $\hat{\nabla}^2\psi^\text{(ext)}=0$, which is a linear combination of solutions $Y_{lm}(\theta,\phi)\hat{r}^{-(l+1)}$. The boundary condition at large radii given in equation~(\ref{eq:combbound}) requires the addition of the constant term $\alpha= aH_0$, and we have defined $\lambda=\beta_{00}Y_{00}$. We expand the constants $\alpha$, $\lambda$ and $\beta_{lm}$ as asymptotic series in $\epsilon$, and functions $T$ and $C$ in spherical harmonics, to give the external solution up to first order in $\epsilon$,
\begin{align}
    \psi^\text{(ext)}(\hat{\mathbfit{r}},t&)=\alpha_0-\frac{\lambda_0}{\hat{r}}+\bigg\{\alpha_1-\frac{\lambda_1}{\hat{r}}-\frac{T_{00}(\hat{r})+\zeta C_{00}(\hat{r})}{\sqrt{4\pi}}\notag\\&-\sum_{l=1}^{\infty}\sum_{m=-l}^l\left[\frac{a_{lm}}{\hat{r}^{l+1}}+T_{lm}(\hat{r})+\zeta C_{lm}(\hat{r})\right]Y_{lm}(\theta,\phi)\bigg\}\epsilon. \label{eq:external}
\end{align}
The constants $\alpha_0$, $\alpha_1$, $\lambda_0$, $\lambda_1$ and $a_{lm}$ are undetermined prior to asymptotic matching. Only the monopole ($l=0$) and quadrupole ($l=2$) coefficients
are non-vanishing in the expansions of $T$ and $C$, with positive, even values of $m$ for $T$ and only $m=0$ for $C$,
\begin{equation}
T_{00}(\hat{r})=-3\sqrt{\pi}(\nu-1)\hat{r}^2, \label{eq:T00}
\end{equation}
\begin{equation}
T_{20}(\hat{r})=3\sqrt{\frac{\pi}{5}}(2+\nu)\hat{r}^2,
\end{equation}
\begin{equation}
T_{22}(\hat{r})=-3\sqrt{\frac{3\pi}{5}}\nu\hat{r}^2,
\end{equation}
\begin{equation}
C_{00}(\hat{r})=-6\sqrt{\pi}\hat{r}^2, \label{eq:C00}
\end{equation}
\begin{equation}
C_{20}(\hat{r})=6\sqrt{\frac{\pi}{5}}\hat{r}^2. \label{eq:C20}
\end{equation}

\subsection{Boundary layer} \label{sec:boundary}
We define the boundary layer as the region in which $\hat{r}_\text{tr}-\hat{r}=\mathcal{O}(\epsilon)$ (see Appendix~\ref{sec:Aboundary}), where the truncation radius $\hat{r}_\text{tr}=\hat{r}_\text{tr}(\Psi)$ defines the spherical boundary of the zeroth-order solution: $\psi_0(\hat{r}_\text{tr})=0$. In order to construct a perturbation expansion that maintains proper ordering in the boundary layer, we scale the radial coordinate and solution with respect to $\epsilon$,
\begin{equation}
    \eta = \frac{\hat{r}_\text{tr}-\hat{r}}{\epsilon}, \label{eq:eta}
\end{equation}
\begin{equation}
\tau = \frac{\psi^\text{(lay)}}{\epsilon}. \label{eq:tau}
\end{equation}
Using the asymptotic expansion of the lower incomplete gamma function \cite[][\href{https://dlmf.nist.gov/8.11.E4}{(8.11.4)}]{NIST:DLMF}, we have $\hat{\rho}(\epsilon\tau)=\mathcal{O}(\epsilon^{5/2})$. Therefore, to first order in $\epsilon$, the source term that differentiates the Poisson and Laplace equations does not appear in the boundary layer, leaving a single differential equation,
\begin{equation}
    \frac{\partial^2\tau}{\partial\eta^2}-\frac{2\epsilon}{\hat{r}_{tr}-\eta\epsilon}\frac{\partial\tau}{\partial\eta}+\frac{\epsilon^2}{(\hat{r}_{tr}-\eta\epsilon)^2}\Lambda^2\tau= -9\epsilon^2\left(1-\nu-2\zeta\right), \label{eq:pl}
\end{equation}    
where $\Lambda^2$ denotes the angular part of the Laplacian in spherical polar coordinates. We expand the boundary layer solution up to first order in $\epsilon$,
\begin{equation}
    \tau=\tau_0+\tau_1\epsilon, \label{eq:tau_expansion}
\end{equation}
and insert this into equation~(\ref{eq:pl}). By collecting same-order terms we obtain a pair of differential equations (see Appendix~\ref{sec:Aboundary}) that integrate directly to
\begin{equation}\tau_0=F_0(\theta,\phi)\eta+G_0(\theta,\phi), \label{eq:tau0}
\end{equation}
\begin{equation}
    \tau_1=\frac{F_0(\theta,\phi)}{\hat{r}_\text{tr}}\eta^2+F_1(\theta,\phi)\eta+G_1(\theta,\phi). \label{eq:tau1}
\end{equation}
The angular functions $F_0$, $G_0$, $F_1$ and $G_1$ are determined via asymptotic matching. Note that since $\zeta$ only appears in equation~(\ref{eq:pl}) at second order in $\epsilon$, the form of the boundary layer description is independent of the presence or absence of asynchronicity.

\subsection{Asymptotic matching} \label{sec:matching}
In order to match the boundary layer solution with the internal and external solutions, we first combine equations~(\ref{eq:tau_expansion})--(\ref{eq:tau1}) to obtain the unscaled first order expansion,
\begin{equation}
    \psi^\text{(lay)}(\hat{\mathbfit{r}},t)=F_0(\theta,\phi)(\hat{r}_\text{tr}-\hat{r})+\big[G_0(\theta,\phi)+F_1(\theta,\phi)(\hat{r}_\text{tr}-\hat{r})\big]\epsilon. \label{eq:layer}
\end{equation}
Comparison with the internal solution Taylor expanded about $\hat
{r}=\hat{r}_\text{tr}$ allows the free angular functions to be determined,
\begin{equation}
    F_0(\theta,\phi)=-\psi_0^{\text{(int)}\prime}(\hat{r}_\text{tr}), \label{eq:F0}
\end{equation}
\begin{equation}
    G_0(\theta,\phi)=\psi_1^\text{(int)}(\hat{r}_\text{tr},\theta,\phi), \label{eq:G0}
\end{equation}
\begin{equation}
    F_1(\theta,\phi)=-\frac{\partial\psi_1^\text{(int)}}{\partial\hat{r}}(\hat{r}_\text{tr},\theta,\phi). \label{eq:F1}
\end{equation}
The external solution is similarly matched with the boundary layer solution, indirectly connecting the internal and external solutions via the angular functions (see Appendix~\ref{sec:Amatching} for details). In this way, the unknown constants in equations~(\ref{eq:internal}) and (\ref{eq:external}) are determined,

\begin{equation}
    \alpha_0=\frac{\lambda_0}{\hat{r}_\text{tr}}, \label{eq:alpha0}
\end{equation}
\begin{equation}
\lambda_0=\hat{r}_\text{tr}^2\psi_0^{\text{(int)}\prime}(\hat{r}_\text{tr}), \label{eq:lambda0}
\end{equation}
\begin{equation}
    \alpha_1=f_{00}(\hat{r}_\text{tr})+f_{00}'(\hat{r}_\text{tr})\hat{r}_\text{tr}+\frac{3\left[T_{00}(\hat{r}_\text{tr})+\zeta C_{00}(\hat{r}_\text{tr})\right]}{\sqrt{4\pi}}, \label{eq:alpha1}
\end{equation}
\begin{equation}
    \lambda_1=f_{00}'(\hat{r}_\text{tr})\hat{r}_\text{tr}^2+\frac{T_{00}(\hat{r}_\text{tr})+\zeta C_{00}(\hat{r}_\text{tr})}{\sqrt{\pi}}\hat{r}_\text{tr}, \label{eq:lambda1}
\end{equation}
\begin{equation}
    a_{20}=-\hat{r}_\text{tr}^3\left[A_{20}\mathcal{\gamma}_2(\hat{r}_\text{tr})+T_{20}(\hat{r}_\text{tr})+\zeta C_{20}(\hat{r}_\text{tr})\right], \label{eq:a20}
\end{equation}
\begin{equation}
    a_{22}=-\hat{r}_\text{tr}^3\left[A_{22}\mathcal{\gamma}_2(\hat{r}_\text{tr})+T_{22}(\hat{r}_\text{tr})\right],
\end{equation}
\begin{equation}
    A_{20}=-\frac{5\left[T_{20}(\hat{r}_\text{tr})+\zeta C_{20}(\hat{r}_\text{tr})\right]}{\mathcal{\gamma}_2'(\hat{r}_\text{tr})\hat{r}_\text{tr}+3\mathcal{\gamma}_2(\hat{r}_\text{tr})},
\end{equation}
\begin{equation}
    A_{22}=-\frac{5T_{22}(\hat{r}_\text{tr})}{\mathcal{\gamma}_2'(\hat{r}_\text{tr})\hat{r}_\text{tr}+3\mathcal{\gamma}_2(\hat{r}_\text{tr})}. \label{eq:A22}
\end{equation}
It also results that $A_{lm}=a_{lm}=0$ for $l\neq2$
and $A_{2m}=a_{2m}=0$ for $m\notin\{0,2\}$, i.e., to first order in $\epsilon$, the unperturbed, \cite{King} model solution is modified by monopole and quadrupole terms only.

By piecewise combination of the internal and external solutions, we now have the form of the global solution $\psi(\hat{\mathbfit{r}},t;\epsilon)$, up to first order in $\epsilon$. The family of models defined by the parameter set $(\Psi,\zeta,\epsilon)$ can therefore be constructed by numerical integration of equations~(\ref{eq:kingpoisson}), (\ref{eq:ode1}) and (\ref{eq:ode2}), the details of which are given in Appendix~\ref{sec:BTIRO}.

\section{Selected properties of the models} \label{sec:properties}
\subsection{Intrinsic radial profiles} \label{sec:profiles}
\begin{figure}
	\includegraphics[width=\columnwidth]{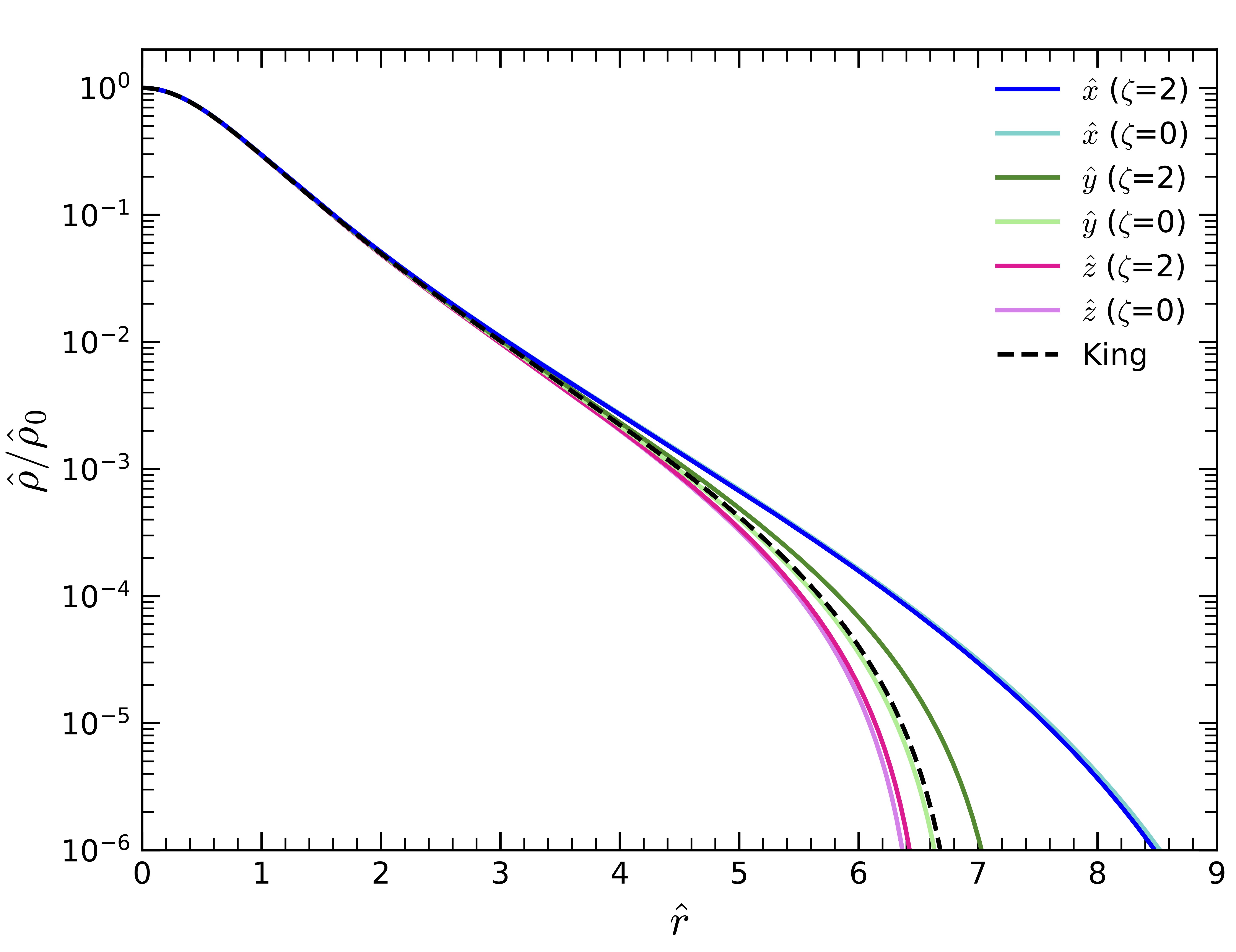}
    \caption{Intrinsic density profiles (normalized to the central value) along the $\hat{x}$-, $\hat{y}$- and $\hat{z}$-axes for a critical first-order model with $\Psi=4$, $\zeta=2$ and $\epsilon=1.211\times10^{-4}$. Also shown are profiles for the corresponding critical synchronous ($\zeta=0$, $\epsilon=1.972\times10^{-4}$) and unperturbed \citet{King} models. The galactic potential is Keplerian ($\nu=3$) in all cases. Note the pale blue $\hat{x}$-axis profile of the synchronous model, the extension of which beyond the corresponding asynchronous profile is narrowly distinguishable.} \label{fig:density}
\end{figure}
\begin{figure}
	\includegraphics[width=\columnwidth]{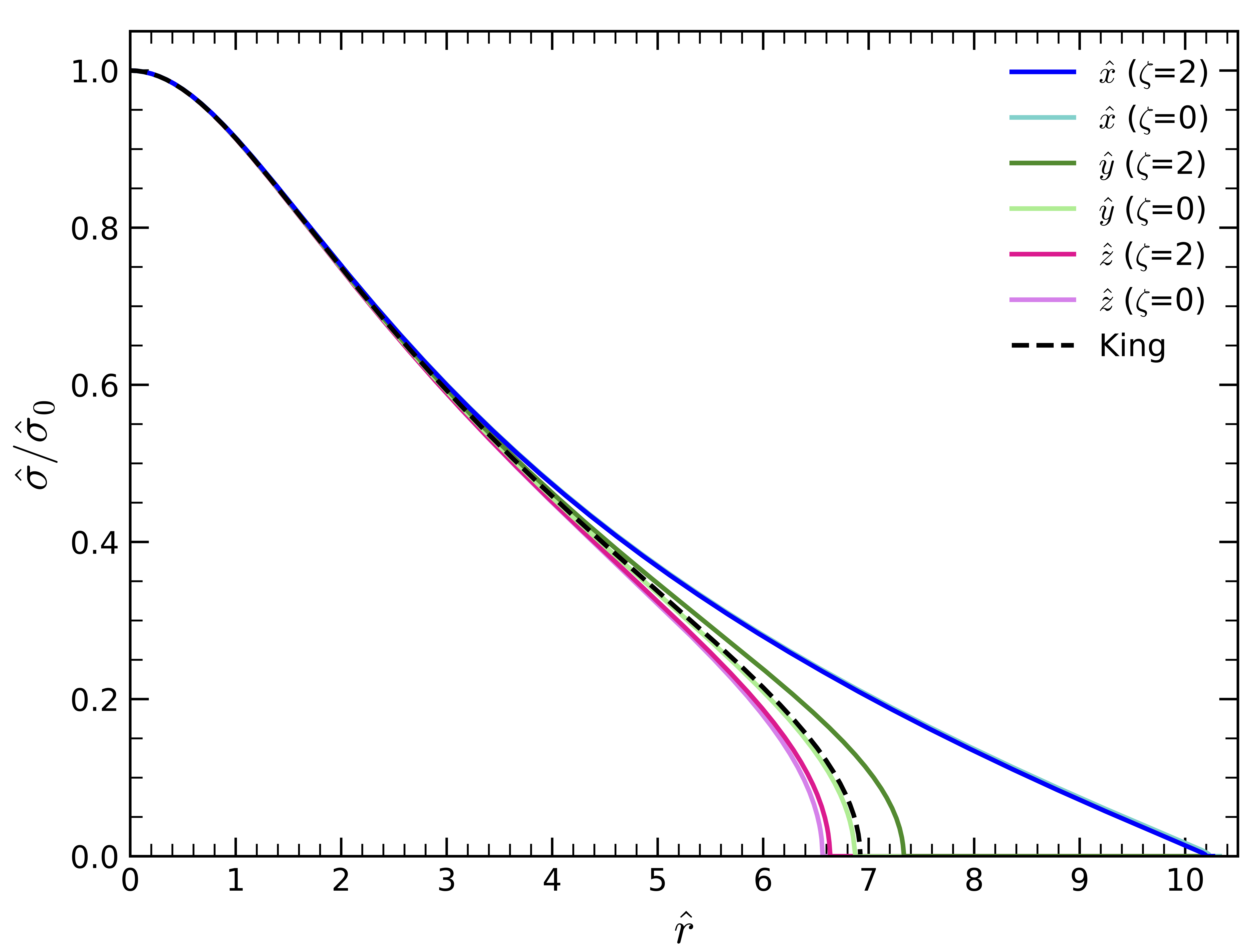}
    \caption{Intrinsic velocity dispersion profiles (normalized to the central value) along the $\hat{x}$-, $\hat{y}$- and $\hat{z}$-axes for the same models as in Fig.~\ref{fig:density}. The $\hat{x}$-axis profiles are almost coincident, with the pale blue line of the synchronous profile just discernible beyond that of the asynchronous model.} \label{fig:dispersion}
\end{figure}
\begin{figure}
	\includegraphics[width=\columnwidth]{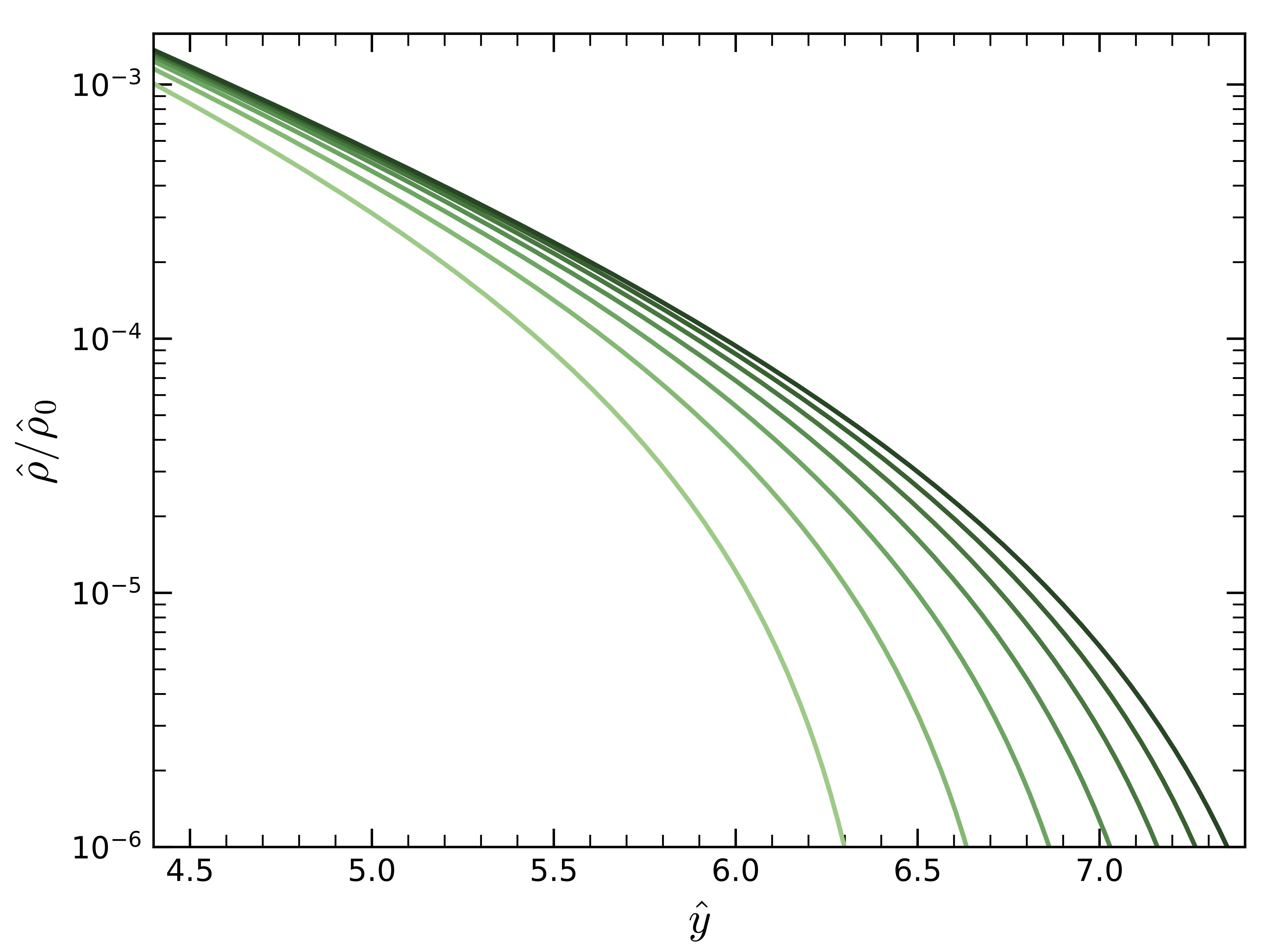}
    \caption{Intrinsic density profiles (normalized to the central value) along the $\hat{y}$-axis for critical first-order models with $\Psi=4$ and $\zeta\in\{-1,0,1,2,3,4,5\}$ (from left to right). The galactic potential is Keplerian ($\nu=3$).} \label{fig:criticaly}
\end{figure}

The models thus constructed exhibit the expected triaxial structure, with reflection symmetry about the three coordinate planes in $(\hat{x},\hat{y},\hat{z})$. In Fig.~\ref{fig:density}, we present the intrinsic density profiles along the $\hat{x}$-, $\hat{y}$-, and $\hat{z}$-axes for a critical model with $\Psi=4$ and $\zeta=2$, together with those for the corresponding (i.e., with the same value of $\Psi$) spherical \cite{King} and critical synchronous models. With respect to the \cite{King} model, there is significant elongation along the $\hat{x}$-axis (the direction radial to the centre of the host galaxy), less pronounced elongation along the $\hat{y}$-axis (the direction of the orbital motion)\footnote{Synchronous models exhibit a slight compression along this axis, on to which increased (decreased) internal rotation superimposes an axisymmetric expansion (compression) in the orbital plane. Consequently, elongation is observed along the $\hat{y}$-axis for $\zeta$ larger than some small positive value that depends weakly on $(\Psi,\epsilon)$, with compression otherwise.}, and compression along the $\hat{z}$-axis (the axis of internal rotation). The same pattern of elongation and compression is apparent in the intrinsic velocity dispersion profiles in Fig.~\ref{fig:dispersion}.

When comparing critical models, it should be recalled that any increase in $\zeta$, at fixed concentration $\Psi$, is accompanied by a decrease in $\epsilon_\text{cr}$ (as shown in Fig.~\ref{fig:parameter1}--\ref{fig:parameter2}). Although the rotational strength parameter $\chi$, which depends on $\zeta$ and $\epsilon$ via equation~(\ref{eq:zeta}), increases, this larger rotational perturbation is combined with a reduction in the tidal perturbation. In Fig.~\ref{fig:density} and Fig.~\ref{fig:dispersion}, comparison with the synchronous ($\zeta=0$) case shows that critical models with increased internal rotation, as parametrised by positive $\zeta$, display greater extension along the $\hat{y}$-axis, but a small decrease in the extension and compression along the $\hat{x}$- and $\hat{z}$-axes, respectively. These effects are found to increase in magnitude for larger positive values of $\zeta$ and to be reversed for $\zeta<0$, as shown in  Fig.~\ref{fig:criticaly}. Deviations from the synchronous $\hat{y}$- and $\hat{z}$-axis profiles emerge only at densities around $\log(\hat{\rho}/\hat{\rho}_0)=-3$, affecting the outer envelope but not the inner regions, where all profiles converge towards that of the relevant \cite{King} model. The effect of $\zeta$ on the $\hat{x}$-axis profile is not only smaller in magnitude, but also arises at even lower densities, around $\log(\hat{\rho}/\hat{\rho}_0)=-5$.

\subsection{Intrinsic sections}
\begin{figure*}
\centering
\begin{subfigure}[b]{0.3\textwidth}
    \includegraphics[width=\textwidth]{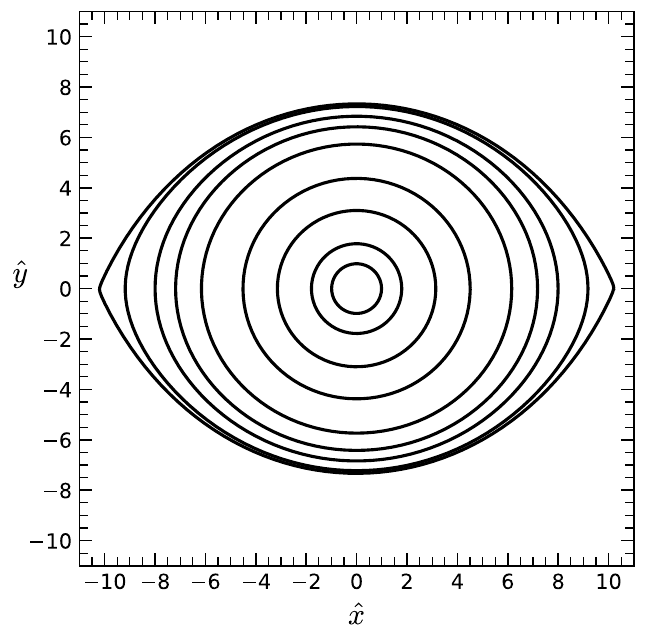}
 \end{subfigure}
 \hfill
 \begin{subfigure}[b]{0.3\textwidth}
     \includegraphics[width=\textwidth]{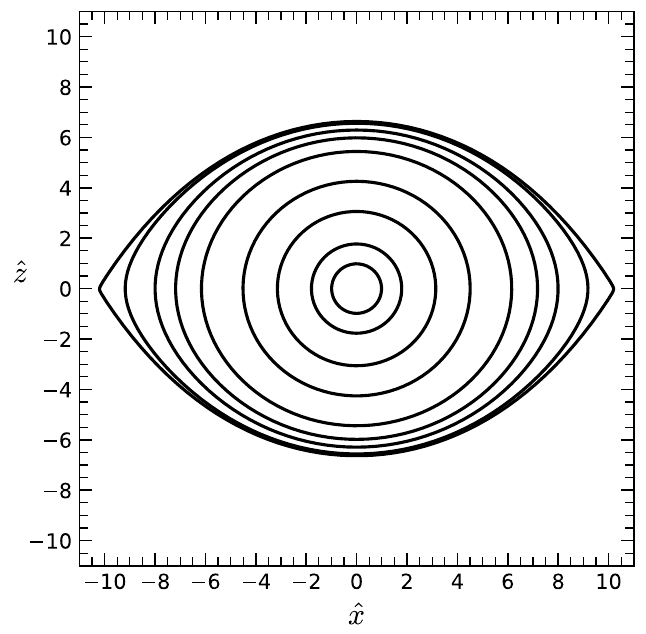}
 \end{subfigure}
 \hfill
 \begin{subfigure}[b]{0.3\textwidth}
    \includegraphics[width=\textwidth]{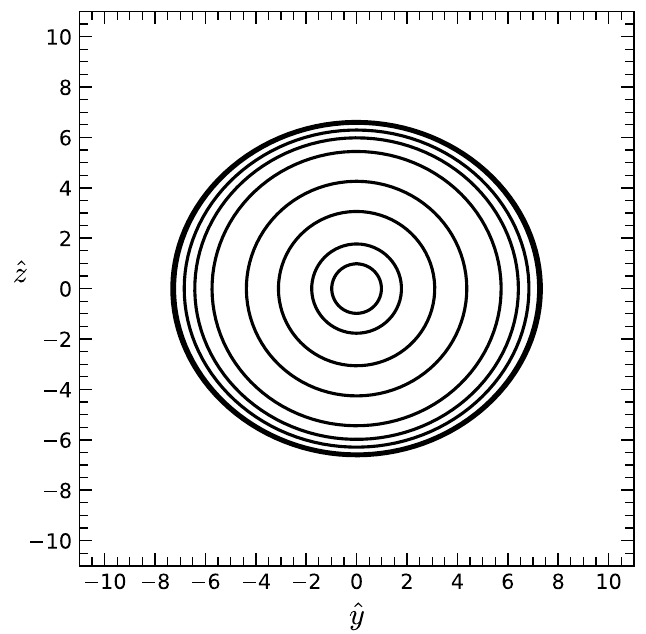}
 \end{subfigure}
\caption{Sections, in the three coordinate planes, of selected equipotential surfaces of a critical first-order model with $\Psi=4$, $\zeta=2$ and 
$\epsilon=1.211\times10^{-4}$. The contours correspond to dimensionless escape energy $\psi\in\{0,0.01,0.05,0.1,0.2,0.5,1,2,3\}$, where $\psi=0$ gives the critical boundary surface. The galactic potential is Keplerian ($\nu=3$).} \label{fig:equipotentials}
\end{figure*}

\begin{figure*}
\centering
\begin{subfigure}[b]{0.3\textwidth}
        \includegraphics[width=\textwidth]{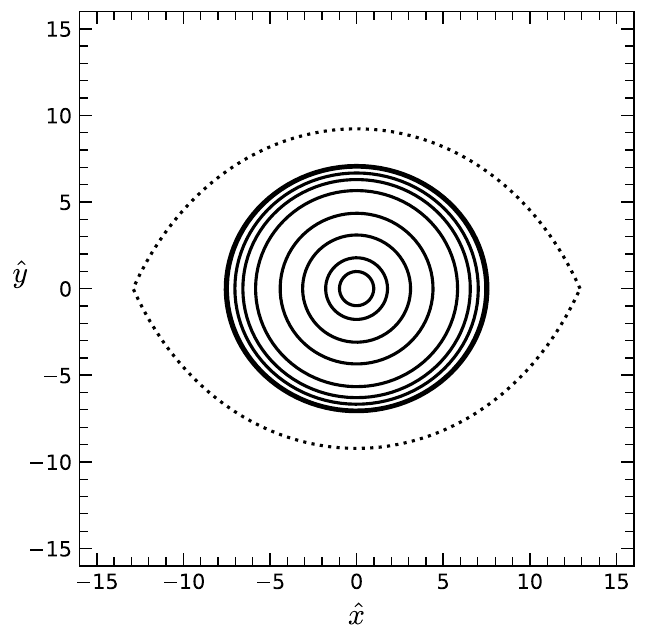}
 \end{subfigure}
 \hfill
 \begin{subfigure}[b]{0.3\textwidth}
     \includegraphics[width=\textwidth]{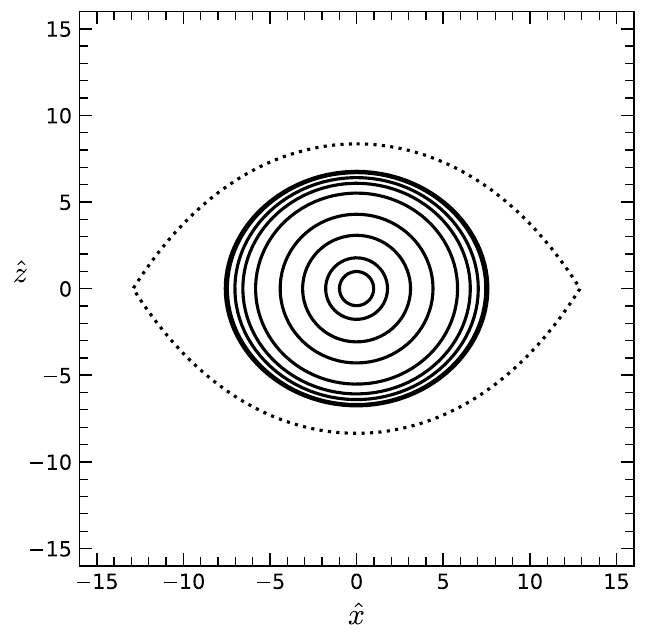}
 \end{subfigure}
 \hfill
 \begin{subfigure}[b]{0.3\textwidth}
\includegraphics[width=\textwidth]{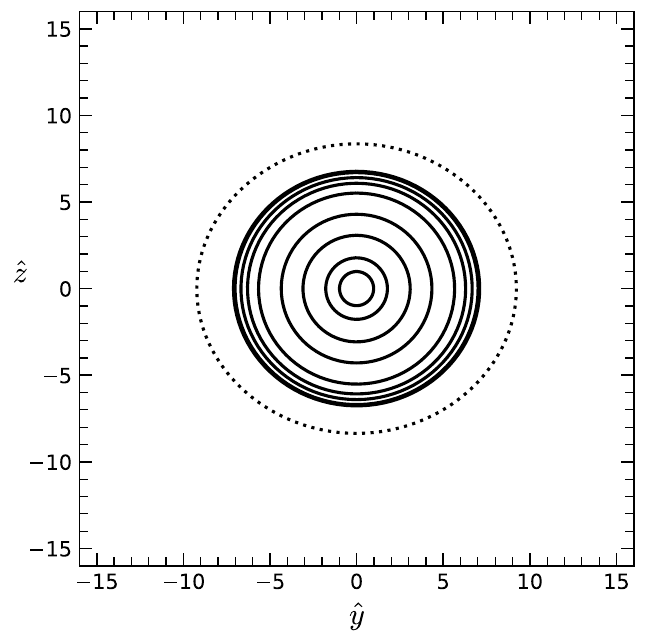}
 \end{subfigure}
\caption{Sections, in the three coordinate planes, of selected equipotential surfaces of a sub-critical (underfilled) first-order model with $\Psi=4$, $\zeta=2$ and $\epsilon=0.5\epsilon_\text{cr}$. The solid contours correspond to values of the dimensionless escape energy between $\psi=0$ and $\psi=3$, as given in Fig.~\ref{fig:equipotentials}. For reference, we also illustrate the critical contour that contains the saddle points (dotted line). The galactic potential is Keplerian ($\nu=3$).}  \label{fig:underfilled}
\end{figure*}

To further characterise the morphology of the models, in Fig.~\ref{fig:equipotentials} we show sections through equipotential surfaces of a critical model with $\Psi=4$ and $\zeta=2$. Since both the density and velocity dispersion are monotonic functions of $\psi$ only (see equations~\ref{eq:density} and \ref{eq:dispersion}), the equipotential surfaces correspond one-to-one with the isodensity and isobaric surfaces. While the inner regions are relatively unaffected by the combined tidal-rotational perturbation, deviating only slightly from spherical symmetry, the outer surfaces exhibit the strongly triaxial structure characteristic of critically perturbed models. For higher values of $\zeta$ (increased asynchronous rotation), the overall deviation from spherical symmetry is reduced in the $\hat{x}\hat{y}$- and $\hat{x}\hat{z}$-planes as the radial profiles along the corresponding pairs of axes move closer together (see Section~\ref{sec:profiles}). The saddle points, located at the intersection of the critical boundary surface with the $\hat{x}$-axis, move closer to the origin with increasing $\zeta$.

Sections through the equipotential surfaces of a sub-critical (underfilled) model with $(\Psi,\zeta,\epsilon)=(4,-1,0.5\epsilon_\text{cr})$ are shown in Fig.~\ref{fig:underfilled}, along with the corresponding critical surface. The equipotential surfaces appear quasi-spherical throughout the model, indicating that its internal structure is largely unaffected by the perturbations. The critical surface, which lies substantially beyond the cluster boundary, does, however, exhibit a triaxial geometry. 

\section{Discussion and Conclusions} \label{sec:conclusions}

A three-parameter family of self-consistent equilibrium dynamical models has been constructed to provide an idealised description of the internal structure and kinematics of a quasi-relaxed stellar system rotating asynchronously within a tidal field. The distribution function that defines these models is derived from that of the spherical, non-rotating \cite{King} models by substituting the relevant Jacobi integral for the single-star energy. A global solution to the associated Poisson-Lagrange equations, across an unknown three-dimensional boundary surface and nearby singularity, is found by the method of matched asymptotic expansion. This is made possible by defining a new asynchronicity parameter that couples the tidal and rotational perturbations, thereby reducing the double perturbation problem to a single expansion in the tidal strength parameter.

For positive values of the asynchronicity parameter, the constructed equilibria have rigid internal rotation of angular frequency greater than that of their orbital motion. Conversely, models with negative asynchronicity rotate more slowly than they orbit around the host galaxy. In the case where the asynchronicity parameter vanishes, the models reduce to the synchronous tidal models of \cite{VarriBertin1} (for the critical case, see also \citealt{HeggieRamamani}). If the tidal strength parameter is set to zero, the one-parameter family of unperturbed \cite{King} models is recovered. By construction, all models are characterised by an isotropic velocity dispersion.

Reflecting the properties of the perturbation potential, asynchronous equilibria have a triaxial structure. Relative to a spherical \cite{King} model of like concentration, we observe extension in the direction radial to the galactic centre, compression along the axis of rotation, and either extension or compression in the direction of the orbital motion (depending on parameter values). At fixed concentration and tidal strength parameter, this aspherical distortion increases in magnitude with the value of the asynchronicity parameter.

Within the relevant three-dimensional parameter space, we can identify two perturbation regimes, distinguished by their proximity to the maximally-deformed critical models. First, there are the close-to-critical models, whose extension in the $x$-direction is close to the maximum allowed for a bound model, and that possess a distinctly triaxial geometry in their outer regions. Second are the sub-critical models, which significantly underfill the volume enclosed by the outermost closed equipotential and exhibit only modest deviations from spherical symmetry. Since the critical value of the tidal strength parameter decreases with increasing asynchronicity, the total parameter space accessible to bound models is diminished. In consequence of this, the aspherical distortion of critical models is actually slightly smaller at higher values of the asynchronicity parameter.

In this article, we have presented full solutions up to first order in the tidal strength parameter. In principle, solutions to arbitrary order may be obtained using the method set out in \cite{VarriBertin1}. However, the benefit of doing so while remaining within the framework of the Hill's approximation, under which we linearised the gradient of the galactic potential, is marginal.

While we have made explicit reference to \cite{King} as the underlying spherical, non-rotating model used, the same procedure is applicable to other truncated isothermal spheres \citep[e.g.,][taking the isotropic, non-rotating limit where appropriate]{Woolley,Prendergast,Wilson,Davoust,LIMEPY}. In fact, ergodic distribution functions that arise from entirely different equations of state, such as low-n polytropes, can be similarly extended. Since the functional form of the perturbation is unchanged, the general structure of the extended models would be conserved. It is the finer details, such as critical parameter values, that would depend on the unperturbed equilibrium used.

Generalisation of the models constructed here to describe stellar systems with asynchronous rotation within a misaligned architecture (i.e., where the rotational and orbital axes are not aligned) is possible by extension of the same method of solution. In a misaligned configuration, the rotational perturbation would break the triaxial symmetry of the tidal perturbation. The multipole expansion, which is truncated at the quadrupolar terms for the aligned models, would necessarily include higher-degree spherical harmonics, even in the first-order solution.

By presenting this class of self-consistent nonspherical equilibrium models characterised by a more realistic combination of physical ingredients, we hope to offer a useful description of tidally-perturbed, rotating stellar systems of interest to both numerical and observational contexts. The relevant distribution function may be sampled for the initialisation of N-body simulations, providing an equilibrium configuration that is consistent with the coexistence of external tides and internal rotation. A repository containing the model solver, with outputs provided in dimensionless, N-body or physical units, is publicly available (see Appendix~\ref{sec:BTIRO}). The intrinsic properties obtained are applicable for use with simulation datasets. Projected observables, although not presented here, may also be straightforwardly derived (see, e.g., \citealt{VarriBertin2}). This would enable model-fitting to photometric, astrometric, and spectroscopic data from Galactic globular clusters with observed rotation profiles and appropriate tidal signatures \citep[see, e.g., the case of 47 Tucanae, as characterised by][]{Bellini}.

Crucially, these models allow evolutionary processes that depend on the presence of an external tidal field, such as the evaporative formation of tidal tails, to be studied in the context of a rotating star cluster. Contemporary numerical studies of tidal tail formation are typically rather crude in their treatment of the phase space of the progenitor system \citep[for a recent review, see][]{Bonaca}, and the underlying fundamental theory has been largely developed under the assumption of simple spherical models \citep[see, e.g.,][]{FukushigeHeggie,Kupper,Just}. In providing a class of equilibria in which internal rotation is incorporated into a tidally perturbed system, an important step is made towards exploring the impact of the progenitor's angular momentum on the properties of the resulting stellar streams \citep[see, e.g., the case of Omega Centauri, as evidenced by][]{Ibata}.

\section*{Acknowledgements}
LAZA acknowledges receipt of a Science and Technology Facilities Council (STFC) PhD studentship (Ref: 2922765). Both authors are grateful for support from a UKRI Future Leaders Fellowship (MR/S018859/1; MR/X011097/1). For the purpose of open access, the authors have applied a Creative Commons Attribution (CC BY) licence to any Author Accepted Manuscript version arising from this submission.

\section*{Data Availability}

Details are provided in Appendix B. 



\bibliographystyle{mnras}
\bibliography{references} 

@ARTICLE{Limber,
       author = {{Limber}, D. Nelson},
        title = "{Surface Forms and Mass Loss for the Components of Close Binaries-General Case of Non-Synchronous Rotation.}",
      journal = {\apj},
         year = 1963,
        month = nov,
       volume = {138},
        pages = {1112},
          doi = {10.1086/147711},
       adsurl = {https://ui.adsabs.harvard.edu/abs/1963ApJ...138.1112L}
}

@ARTICLE{Kopal,
       author = {{Kopal}, Z.},
        title = "{Evolutionary processes in close binary stars}",
      journal = {Ann. Astrophys.},
         year = 1956,
        month = jan,
       volume = {19},
        pages = {298},
       adsurl = {https://ui.adsabs.harvard.edu/abs/1956AnAp...19..298K}
}

@ARTICLE{Smith,
       author = {{Smith}, B.~L.},
        title = "{Rapidly Rotating Polytropes as a Net of Singular Perturbation Problems}",
      journal = {\apss},
         year = 1975,
        month = jul,
       volume = {35},
       number = {2},
        pages = {223-240},
          doi = {10.1007/BF00636993},
       adsurl = {https://ui.adsabs.harvard.edu/abs/1975Ap&SS..35..223S}
}

@ARTICLE{King,
       author = {{King}, Ivan R.},
        title = "{The structure of star clusters. III. Some simple dynamical models}",
      journal = {\aj},
         year = 1966,
        month = feb,
       volume = {71},
        pages = {64},
          doi = {10.1086/109857},
       adsurl = {https://ui.adsabs.harvard.edu/abs/1966AJ.....71...64K}
}

@book{HeggieHut,
  author    = {Heggie, D. and Hut, P.}, 
  title     = {The Gravitational Million–Body Problem},
  publisher = {Cambridge Univ. Press},
  year      = {2003},
  address   = {Cambridge},
}

@book{VanDyke,
  author    = {Van Dyke, M.}, 
  title     = {Perturbation Methods in Fluid Mechanics},
  publisher = {Parabolic Press},
  year      = {1975},
  address   = {Stanford},
}

@ARTICLE{VarriBertin1,
       author = {{Bertin}, G. and {Varri}, A.~L.},
        title = "{The Construction of Nonspherical Models of Quasi-Relaxed Stellar Systems}",
      journal = {\apj},
         year = 2008,
        month = dec,
       volume = {689},
       number = {2},
        pages = {1005-1019},
          doi = {10.1086/592684},
archivePrefix = {arXiv},
       eprint = {0808.2432},
 primaryClass = {astro-ph},
       adsurl = {https://ui.adsabs.harvard.edu/abs/2008ApJ...689.1005B}
}

@ARTICLE{VarriBertin2,
       author = {{Varri}, A.~L. and {Bertin}, G.},
        title = "{Properties of Quasi-relaxed Stellar Systems in an External Tidal Field}",
      journal = {\apj},
         year = 2009,
        month = oct,
       volume = {703},
       number = {2},
        pages = {1911-1922},
          doi = {10.1088/0004-637X/703/2/1911},
archivePrefix = {arXiv},
       eprint = {0908.2388},
 primaryClass = {astro-ph.GA},
       adsurl = {https://ui.adsabs.harvard.edu/abs/2009ApJ...703.1911V}
}

@ARTICLE{VarriBertin3,
       author = {{Varri}, A.~L. and {Bertin}, G.},
        title = "{Self-consistent models of quasi-relaxed rotating stellar systems}",
      journal = {\aap},
         year = 2012,
        month = apr,
       volume = {540},
          eid = {A94},
        pages = {A94},
          doi = {10.1051/0004-6361/201118300},
archivePrefix = {arXiv},
       eprint = {1201.1899},
 primaryClass = {astro-ph.GA},
       adsurl = {https://ui.adsabs.harvard.edu/abs/2012A&A...540A..94V}
}

@ARTICLE{HeggieRamamani,
       author = {{Heggie}, D.~C. and {Ramamani}, N.},
        title = "{Approximate self-consistent models for tidally truncated star clusters}",
      journal = {\mnras},
         year = 1995,
        month = jan,
       volume = {272},
       number = {2},
        pages = {317-322},
          doi = {10.1093/mnras/272.2.317},
       adsurl = {https://ui.adsabs.harvard.edu/abs/1995MNRAS.272..317H}
}

@ARTICLE{Woolley,
       author = {{Woolley}, R.~V.~D.~R.},
        title = "{A study of the equilibrium of globular clusters}",
      journal = {\mnras},
         year = 1954,
        month = jan,
       volume = {114},
        pages = {191},
          doi = {10.1093/mnras/114.2.191},
       adsurl = {https://ui.adsabs.harvard.edu/abs/1954MNRAS.114..191W}
}

@ARTICLE{Wilson,
       author = {{Wilson}, C.~P.},
        title = "{Dynamical models of elliptical galaxies.}",
      journal = {\aj},
         year = 1975,
        month = mar,
       volume = {80},
        pages = {175-187},
          doi = {10.1086/111729},
       adsurl = {https://ui.adsabs.harvard.edu/abs/1975AJ.....80..175W}
}

@ARTICLE{Prendergast,
       author = {{Prendergast}, Kevin H. and {Tomer}, Eugene},
        title = "{Self-Consistent Models of Elliptical Galaxies}",
      journal = {\aj},
         year = 1970,
        month = aug,
       volume = {75},
        pages = {674},
          doi = {10.1086/111008},
       adsurl = {https://ui.adsabs.harvard.edu/abs/1970AJ.....75..674P}
}

@ARTICLE{Davoust,
       author = {{Davoust}, E.},
        title = "{Analytical models for spherical stellar systems.}",
      journal = {\aap},
         year = 1977,
        month = nov,
       volume = {61},
       number = {3},
        pages = {391-396},
       adsurl = {https://ui.adsabs.harvard.edu/abs/1977A&A....61..391D}
}

@ARTICLE{LIMEPY,
       author = {{Gieles}, Mark and {Zocchi}, Alice},
        title = "{A family of lowered isothermal models}",
      journal = {\mnras},
         year = 2015,
        month = nov,
       volume = {454},
       number = {1},
        pages = {576-592},
          doi = {10.1093/mnras/stv1848},
archivePrefix = {arXiv},
       eprint = {1508.02120},
 primaryClass = {astro-ph.IM},
       adsurl = {https://ui.adsabs.harvard.edu/abs/2015MNRAS.454..576G}
}

@ARTICLE{Kupper,
       author = {{K{\"u}pper}, Andreas H.~W. and {MacLeod}, Andrew and {Heggie}, Douglas C.},
        title = "{On the structure of tidal tails}",
      journal = {\mnras},
         year = 2008,
        month = jul,
       volume = {387},
       number = {3},
        pages = {1248-1252},
          doi = {10.1111/j.1365-2966.2008.13323.x},
archivePrefix = {arXiv},
       eprint = {0804.2476},
 primaryClass = {astro-ph},
       adsurl = {https://ui.adsabs.harvard.edu/abs/2008MNRAS.387.1248K}
}

@ARTICLE{Just,
       author = {{Just}, A. and {Berczik}, P. and {Petrov}, M.~I. and {Ernst}, A.},
        title = "{Quantitative analysis of clumps in the tidal tails of star clusters}",
      journal = {\mnras},
         year = 2009,
        month = jan,
       volume = {392},
       number = {3},
        pages = {969-981},
          doi = {10.1111/j.1365-2966.2008.14099.x},
archivePrefix = {arXiv},
       eprint = {0808.3293},
 primaryClass = {astro-ph},
       adsurl = {https://ui.adsabs.harvard.edu/abs/2009MNRAS.392..969J}
}

@ARTICLE{Ibata,
       author = {{Ibata}, Rodrigo A. and {Bellazzini}, Michele and {Malhan}, Khyati and {Martin}, Nicolas and {Bianchini}, Paolo},
        title = "{Identification of the long stellar stream of the prototypical massive globular cluster {\ensuremath{\omega}} Centauri}",
      journal = {Nature Astronomy},
         year = 2019,
        month = apr,
       volume = {3},
        pages = {667-672},
          doi = {10.1038/s41550-019-0751-x},
archivePrefix = {arXiv},
       eprint = {1902.09544},
 primaryClass = {astro-ph.GA},
       adsurl = {https://ui.adsabs.harvard.edu/abs/2019NatAs...3..667I}
}

@ARTICLE{Bonaca,
       author = {{Bonaca}, Ana and {Price-Whelan}, Adrian M.},
        title = "{Stellar streams in the Gaia era}",
      journal = {\nar},
         year = 2025,
        month = jun,
       volume = {100},
          eid = {101713},
        pages = {101713},
          doi = {10.1016/j.newar.2024.101713},
archivePrefix = {arXiv},
       eprint = {2405.19410},
 primaryClass = {astro-ph.GA},
       adsurl = {https://ui.adsabs.harvard.edu/abs/2025NewAR.10001713B}
}

@ARTICLE{FukushigeHeggie,
       author = {{Fukushige}, T. and {Heggie}, D.~C.},
        title = "{The time-scale of escape from star clusters}",
      journal = {\mnras},
         year = 2000,
        month = nov,
       volume = {318},
       number = {3},
        pages = {753-761},
          doi = {10.1046/j.1365-8711.2000.03811.x},
archivePrefix = {arXiv},
       eprint = {astro-ph/9910468},
 primaryClass = {astro-ph},
       adsurl = {https://ui.adsabs.harvard.edu/abs/2000MNRAS.318..753F}
}

@ARTICLE{Hill,
    author = {{Hill}, G.~W.},
    title = "{Researches in the Lunar Theory}",
    journal = {American Journal of Mathematics},
    year = 1878,
    volume = {1},
    number = {1},
    pages = {5-26},    
    ISSN = {00029327, 10806377},
    URL = {http://www.jstor.org/stable/2369430},publisher = {Johns Hopkins University Press},
}

@ARTICLE{Gaia,
       author = {{Gaia Collaboration,}},
        title = "{Gaia Data Release 2. Kinematics of globular clusters and dwarf galaxies around the Milky Way}",
      journal = {\aap},
         year = 2018,
        month = aug,
       volume = {616},
          eid = {A12},
        pages = {A12},
          doi = {10.1051/0004-6361/201832698},
archivePrefix = {arXiv},
       eprint = {1804.09381},
 primaryClass = {astro-ph.GA},
       adsurl = {https://ui.adsabs.harvard.edu/abs/2018A&A...616A..12G}
}

@ARTICLE{Chen,
       author = {{Chen}, C.~W. and {Chen}, W.~P.},
        title = "{Morphological Distortion of Galactic Globular Clusters}",
      journal = {\apj},
         year = 2010,
        month = oct,
       volume = {721},
       number = {2},
        pages = {1790-1819},
          doi = {10.1088/0004-637X/721/2/1790},
       adsurl = {https://ui.adsabs.harvard.edu/abs/2010ApJ...721.1790C}
}

@ARTICLE{MultiplePops,
       author = {{Milone}, Antonino P. and {Marino}, Anna F.},
        title = "{Multiple Populations in Star Clusters}",
      journal = {Universe},
         year = 2022,
        month = jun,
       volume = {8},
       number = {7},
          eid = {359},
        pages = {359},
          doi = {10.3390/universe8070359},
archivePrefix = {arXiv},
       eprint = {2206.10564},
 primaryClass = {astro-ph.GA},
       adsurl = {https://ui.adsabs.harvard.edu/abs/2022Univ....8..359M}
}

@misc{NIST:DLMF,
         author = "{\relax DLMF}",
       title = "{NIST Digital Library of Mathematical Functions}",
    year = 2025,
howpublished = "\url{https://dlmf.nist.gov/}, Release 1.2.4 of 2025-03-15",
        note = "F.~W.~J. Olver, A.~B. {Olde Daalhuis}, D.~W. Lozier, B.~I. Schneider,
                R.~F. Boisvert, C.~W. Clark, B.~R. Miller, B.~V. Saunders,
                H.~S. Cohl, and M.~A. McClain, eds."}

@ARTICLE{Chandra1,
       author = {{Chandrasekhar}, S.},
        title = "{The equilibrium of distorted polytropes. I. The rotational problem}",
      journal = {\mnras},
         year = 1933,
        month = mar,
       volume = {93},
        pages = {390-406},
          doi = {10.1093/mnras/93.5.390},
       adsurl = {https://ui.adsabs.harvard.edu/abs/1933MNRAS..93..390C}
}

@ARTICLE{Chandra2,
       author = {{Chandrasekhar}, S.},
        title = "{The equilibrium of distorted polytropes. II. the tidal problem}",
      journal = {\mnras},
         year = 1933,
        month = apr,
       volume = {93},
        pages = {449},
          doi = {10.1093/mnras/93.6.449},
       adsurl = {https://ui.adsabs.harvard.edu/abs/1933MNRAS..93..449C}
}

@ARTICLE{Chandra3,
       author = {{Chandrasekhar}, S.},
        title = "{The equilibrium of distorted polytropes. III. the double star problem}",
      journal = {\mnras},
         year = 1933,
        month = apr,
       volume = {93},
        pages = {462},
          doi = {10.1093/mnras/93.6.462},
       adsurl = {https://ui.adsabs.harvard.edu/abs/1933MNRAS..93..462C}
}

@ARTICLE{Tiongco,
       author = {{Tiongco}, Maria A. and {Vesperini}, Enrico and {Varri}, Anna Lisa},
        title = "{Kinematical evolution of tidally limited star clusters: the role of retrograde stellar orbits}",
      journal = {\mnras},
         year = 2016,
        month = sep,
       volume = {461},
       number = {1},
        pages = {402-411},
          doi = {10.1093/mnras/stw1341},
archivePrefix = {arXiv},
       eprint = {1606.06743},
 primaryClass = {astro-ph.GA},
       adsurl = {https://ui.adsabs.harvard.edu/abs/2016MNRAS.461..402T}
}

@INPROCEEDINGS{Plavec,
       author = {{Plavec}, M.},
        title = "{49. Dynamical Instability of the Components of Close Binary Systems}",
    booktitle = {Liege International Astrophysical Colloquia},
         year = 1958,
       series = {Liege International Astrophysical Colloquia},
       volume = {8},
        month = jan,
        pages = {411-420},
       adsurl = {https://ui.adsabs.harvard.edu/abs/1958LIACo...8..411P}
}

@ARTICLE{Kruszewski,
       author = {{Kruszewski}, A.},
        title = "{Exchange of Matter in Close Binary Systems. I. Equilibrium Configurations in the Case of Deviations from Synchronism}",
      journal = {\actaa},
         year = 1963,
        month = jan,
       volume = {13},
        pages = {106},
       adsurl = {https://ui.adsabs.harvard.edu/abs/1963AcA....13..106K}
}

@ARTICLE{Lubow,
       author = {{Lubow}, S.~H.},
        title = "{Equilibrium states of nonsynchronous stars in detached binaries.}",
      journal = {\apj},
         year = 1979,
        month = may,
       volume = {229},
        pages = {1008-1022},
          doi = {10.1086/157036},
       adsurl = {https://ui.adsabs.harvard.edu/abs/1979ApJ...229.1008L}
}

@ARTICLE{AvniSchiller,
       author = {{Avni}, Y. and {Schiller}, N.},
        title = "{Generalized Roche potential for misaligned binary systems - Properties of the critical lobe}",
      journal = {\apj},
         year = 1982,
        month = jun,
       volume = {257},
        pages = {703-714},
          doi = {10.1086/160025},
       adsurl = {https://ui.adsabs.harvard.edu/abs/1982ApJ...257..703A}
}

@ARTICLE{Sepinsky,
       author = {{Sepinsky}, J.~F. and {Willems}, B. and {Kalogera}, V.},
        title = "{Equipotential Surfaces and Lagrangian Points in Nonsynchronous, Eccentric Binary and Planetary Systems}",
      journal = {\apj},
         year = 2007,
        month = may,
       volume = {660},
       number = {2},
        pages = {1624-1635},
          doi = {10.1086/513736},
archivePrefix = {arXiv},
       eprint = {astro-ph/0612508},
 primaryClass = {astro-ph},
       adsurl = {https://ui.adsabs.harvard.edu/abs/2007ApJ...660.1624S}
}

@INPROCEEDINGS{NaylorAnand,
       author = {{Naylor}, M.~D.~T. and {Anand}, S.~P.~S.},
        title = "{Structure of Close Binaries. I Polytropes}",
    booktitle = {IAU Colloquium 4: Stellar Rotation},
         year = 1970,
       editor = {{Slettebak}, Arne},
        month = jan,
        pages = {157},
       adsurl = {https://ui.adsabs.harvard.edu/abs/1970stro.coll..157N}
}

@ARTICLE{Naylor,
       author = {{Naylor}, M.~D.~T.},
        title = "{Structure of Close Binaries. IV: Non-Synchronous Polytropes}",
      journal = {\apss},
         year = 1972,
        month = sep,
       volume = {18},
       number = {1},
        pages = {85-88},
          doi = {10.1007/BF00645281},
       adsurl = {https://ui.adsabs.harvard.edu/abs/1972Ap&SS..18...85N}
}

@ARTICLE{Todoran,
       author = {{Todoran}, Ioan},
        title = "{Equipotential Surfaces in Close Binary Systems - Remarks on the Time-Dependent Potential Function}",
      journal = {\apss},
         year = 1992,
        month = jan,
       volume = {187},
       number = {1},
        pages = {119-126},
          doi = {10.1007/BF00642692},
       adsurl = {https://ui.adsabs.harvard.edu/abs/1992Ap&SS.187..119T}
}

@ARTICLE{Vasiliev,
       author = {{Vasiliev}, Eugene and {Baumgardt}, Holger},
        title = "{Gaia EDR3 view on galactic globular clusters}",
      journal = {\mnras},
         year = 2021,
        month = aug,
       volume = {505},
       number = {4},
        pages = {5978-6002},
          doi = {10.1093/mnras/stab1475},
archivePrefix = {arXiv},
       eprint = {2102.09568},
 primaryClass = {astro-ph.GA},
       adsurl = {https://ui.adsabs.harvard.edu/abs/2021MNRAS.505.5978V}
}

@ARTICLE{Bellini,
       author = {{Bellini}, A. and {Bianchini}, P. and {Varri}, A.~L. and {Anderson}, J. and {Piotto}, G. and {van der Marel}, R.~P. and {Vesperini}, E. and {Watkins}, L.~L.},
        title = "{Hubble Space Telescope Proper Motion (HSTPROMO) Catalogs of Galactic Globular Clusters. V. The Rapid Rotation of 47 Tuc Traced and Modeled in Three Dimensions}",
      journal = {\apj},
         year = 2017,
        month = aug,
       volume = {844},
       number = {2},
          eid = {167},
        pages = {167},
          doi = {10.3847/1538-4357/aa7c5f},
archivePrefix = {arXiv},
       eprint = {1706.08974},
 primaryClass = {astro-ph.GA},
       adsurl = {https://ui.adsabs.harvard.edu/abs/2017ApJ...844..167B}
}

@ARTICLE{HST,
       author = {{Libralato}, Mattia and {Bellini}, Andrea and {van der Marel}, Roeland P. and {Anderson}, Jay and {Watkins}, Laura L. and {Piotto}, Giampaolo and {Ferraro}, Francesco R. and {Nardiello}, Domenico and {Vesperini}, Enrico},
        title = "{Hubble Space Telescope Proper Motion (HSTPROMO) Catalogs of Galactic Globular Cluster. VI. Improved Data Reduction and Internal-kinematic Analysis of NGC 362}",
      journal = {\apj},
         year = 2018,
        month = jul,
       volume = {861},
       number = {2},
          eid = {99},
        pages = {99},
          doi = {10.3847/1538-4357/aac6c0},
archivePrefix = {arXiv},
       eprint = {1805.05332},
 primaryClass = {astro-ph.SR},
       adsurl = {https://ui.adsabs.harvard.edu/abs/2018ApJ...861...99L}
}

@ARTICLE{MUSE,
       author = {{Kamann}, S. and {Husser}, T. -O. and {Dreizler}, S. and {Emsellem}, E. and {Weilbacher}, P.~M. and {Martens}, S. and {Bacon}, R. and {den Brok}, M. and {Giesers}, B. and {Krajnovi{\'c}}, D. and {Roth}, M.~M. and {Wendt}, M. and {Wisotzki}, L.},
        title = "{A stellar census in globular clusters with MUSE: The contribution of rotation to cluster dynamics studied with 200 000 stars}",
      journal = {\mnras},
         year = 2018,
        month = feb,
       volume = {473},
       number = {4},
        pages = {5591-5616},
          doi = {10.1093/mnras/stx2719},
archivePrefix = {arXiv},
       eprint = {1710.07257},
 primaryClass = {astro-ph.GA},
       adsurl = {https://ui.adsabs.harvard.edu/abs/2018MNRAS.473.5591K}
}

@ARTICLE{Geyer,
       author = {{Geyer}, E.~H. and {Hopp}, U. and {Nelles}, B.},
        title = "{Ellipticity variations within some globular clusters of the galaxy and the Magellanic Clouds.}",
      journal = {\aap},
         year = 1983,
        month = sep,
       volume = {125},
        pages = {359-367},
       adsurl = {https://ui.adsabs.harvard.edu/abs/1983A&A...125..359G}
}

@ARTICLE{WhiteShawl,
       author = {{White}, Raymond E. and {Shawl}, Stephen J.},
        title = "{Axial Ratios and Orientations for 100 Galactic Globular Star Clusters}",
      journal = {\apj},
         year = 1987,
        month = jun,
       volume = {317},
        pages = {246},
          doi = {10.1086/165273},
       adsurl = {https://ui.adsabs.harvard.edu/abs/1987ApJ...317..246W}
}

@ARTICLE{Bergh,
       author = {{van den Bergh}, Sidney},
        title = "{The Flattening of Globular Clusters}",
      journal = {\aj},
         year = 2008,
        month = may,
       volume = {135},
       number = {5},
        pages = {1731-1737},
          doi = {10.1088/0004-6256/135/5/1731},
archivePrefix = {arXiv},
       eprint = {0802.4061},
 primaryClass = {astro-ph},
       adsurl = {https://ui.adsabs.harvard.edu/abs/2008AJ....135.1731V}
}

@ARTICLE{Sollima,
       author = {{Sollima}, A. and {Baumgardt}, H. and {Hilker}, M.},
        title = "{The eye of Gaia on globular clusters kinematics: internal rotation}",
      journal = {\mnras},
         year = 2019,
        month = may,
       volume = {485},
       number = {1},
        pages = {1460-1476},
          doi = {10.1093/mnras/stz505},
archivePrefix = {arXiv},
       eprint = {1902.05895},
 primaryClass = {astro-ph.GA},
       adsurl = {https://ui.adsabs.harvard.edu/abs/2019MNRAS.485.1460S}
}

@ARTICLE{Zhang,
       author = {{Zhang}, Shumeng and {Mackey}, Dougal and {Da Costa}, Gary S.},
        title = "{A search for stellar structures around nine outer halo globular clusters in the Milky Way}",
      journal = {\mnras},
         year = 2022,
        month = jul,
       volume = {513},
       number = {3},
        pages = {3136-3164},
          doi = {10.1093/mnras/stac751},
archivePrefix = {arXiv},
       eprint = {2111.09072},
 primaryClass = {astro-ph.GA},
       adsurl = {https://ui.adsabs.harvard.edu/abs/2022MNRAS.513.3136Z}
}

@ARTICLE{Dalessandro,
       author = {{Dalessandro}, E. and {Cadelano}, M. and {Della Croce}, A. and {Aros}, F.~I. and {White}, E.~B. and {Vesperini}, E. and {Fanelli}, C. and {Ferraro}, F.~R. and {Lanzoni}, B. and {Leanza}, S. and {Origlia}, L.},
        title = "{A 3D view of multiple populations' kinematics in Galactic globular clusters}",
      journal = {\aap},
         year = 2024,
        month = nov,
       volume = {691},
          eid = {A94},
        pages = {A94},
          doi = {10.1051/0004-6361/202451054},
archivePrefix = {arXiv},
       eprint = {2409.03827},
 primaryClass = {astro-ph.GA},
       adsurl = {https://ui.adsabs.harvard.edu/abs/2024A&A...691A..94D}
}

@ARTICLE{Cordoni,
       author = {{Cordoni}, G. and {Milone}, A.~P. and {Mastrobuono-Battisti}, A. and {Marino}, A.~F. and {Lagioia}, E.~P. and {Tailo}, M. and {Baumgardt}, H. and {Hilker}, M.},
        title = "{Three-component Kinematics of Multiple Stellar Populations in Globular Clusters with Gaia and VLT}",
      journal = {\apj},
         year = 2020,
        month = jan,
       volume = {889},
       number = {1},
          eid = {18},
        pages = {18},
          doi = {10.3847/1538-4357/ab5aee},
archivePrefix = {arXiv},
       eprint = {1905.09908},
 primaryClass = {astro-ph.SR},
       adsurl = {https://ui.adsabs.harvard.edu/abs/2020ApJ...889...18C}
}

@ARTICLE{McLaughlin,
       author = {{McLaughlin}, Dean E. and {van der Marel}, Roeland P.},
        title = "{Resolved Massive Star Clusters in the Milky Way and Its Satellites: Brightness Profiles and a Catalog of Fundamental Parameters}",
      journal = {\apjs},
         year = 2005,
        month = dec,
       volume = {161},
       number = {2},
        pages = {304-360},
          doi = {10.1086/497429},
archivePrefix = {arXiv},
       eprint = {astro-ph/0605132},
 primaryClass = {astro-ph},
       adsurl = {https://ui.adsabs.harvard.edu/abs/2005ApJS..161..304M}
}

@book{Horedt,
  author    = {G. P. Horedt}, 
  title     = {Polytropes: Applications in Astrophysics and Related Fields},
  publisher = {Kluwer Academic Publishers},
  year      = {2004},
  address   = {Dordrecht},
}

@ARTICLE{HST2,
       author = {{Watkins}, Laura L. and {van der Marel}, Roeland P. and {Bellini}, Andrea and {Anderson}, Jay},
        title = "{Hubble Space Telescope Proper Motion (HSTPROMO) Catalogs of Galactic Globular Cluster. II. Kinematic Profiles and Maps}",
      journal = {\apj},
         year = 2015,
        month = apr,
       volume = {803},
       number = {1},
          eid = {29},
        pages = {29},
          doi = {10.1088/0004-637X/803/1/29},
archivePrefix = {arXiv},
       eprint = {1502.00005},
 primaryClass = {astro-ph.GA},
       adsurl = {https://ui.adsabs.harvard.edu/abs/2015ApJ...803...29W}
}

@ARTICLE{Ferraro,
       author = {{Ferraro}, F.~R. and {Mucciarelli}, A. and {Lanzoni}, B. and {Pallanca}, C. and {Lapenna}, E. and {Origlia}, L. and {Dalessandro}, E. and {Valenti}, E. and {Beccari}, G. and {Bellazzini}, M. and {Vesperini}, E. and {Varri}, A. and {Sollima}, A.},
        title = "{MIKiS: The Multi-instrument Kinematic Survey of Galactic Globular Clusters. I. Velocity Dispersion Profiles and Rotation Signals of 11 Globular Clusters}",
      journal = {\apj},
         year = 2018,
        month = jun,
       volume = {860},
       number = {1},
          eid = {50},
        pages = {50},
          doi = {10.3847/1538-4357/aabe2f},
archivePrefix = {arXiv},
       eprint = {1804.08618},
 primaryClass = {astro-ph.GA},
       adsurl = {https://ui.adsabs.harvard.edu/abs/2018ApJ...860...50F}
}

@ARTICLE{Jordi,
       author = {{Jordi}, K. and {Grebel}, E.~K.},
        title = "{Search for extratidal features around 17 globular clusters in the Sloan Digital Sky Survey}",
      journal = {\aap},
         year = 2010,
        month = nov,
       volume = {522},
          eid = {A71},
        pages = {A71},
          doi = {10.1051/0004-6361/201014392},
archivePrefix = {arXiv},
       eprint = {1008.2966},
 primaryClass = {astro-ph.GA},
       adsurl = {https://ui.adsabs.harvard.edu/abs/2010A&A...522A..71J}
}

@ARTICLE{Bianchini,
       author = {{Bianchini}, P. and {van der Marel}, R.~P. and {del Pino}, A. and {Watkins}, L.~L. and {Bellini}, A. and {Fardal}, M.~A. and {Libralato}, M. and {Sills}, A.},
        title = "{The internal rotation of globular clusters revealed by Gaia DR2}",
      journal = {\mnras},
         year = 2018,
        month = dec,
       volume = {481},
       number = {2},
        pages = {2125-2139},
          doi = {10.1093/mnras/sty2365},
archivePrefix = {arXiv},
       eprint = {1806.02580},
 primaryClass = {astro-ph.GA},
       adsurl = {https://ui.adsabs.harvard.edu/abs/2018MNRAS.481.2125B}
}

@ARTICLE{Cordero,
       author = {{Cordero}, M.~J. and {H{\'e}nault-Brunet}, V. and {Pilachowski}, C.~A. and {Balbinot}, E. and {Johnson}, C.~I. and {Varri}, A.~L.},
        title = "{Differences in the rotational properties of multiple stellar populations in M13: a faster rotation for the `extreme' chemical subpopulation}",
      journal = {\mnras},
         year = 2017,
        month = mar,
       volume = {465},
       number = {3},
        pages = {3515-3535},
          doi = {10.1093/mnras/stw2812},
archivePrefix = {arXiv},
       eprint = {1610.09374},
 primaryClass = {astro-ph.GA},
       adsurl = {https://ui.adsabs.harvard.edu/abs/2017MNRAS.465.3515C}
}

@ARTICLE{Leitinger,
       author = {{Leitinger}, E.~I. and {Baumgardt}, H. and {Cabrera-Ziri}, I. and {Hilker}, M. and {Carbajo-Hijarrubia}, J. and {Gieles}, M. and {Husser}, T.~O. and {Kamann}, S.},
        title = "{The kinematics of 30 Milky Way globular clusters and the multiple stellar populations within}",
      journal = {\aap},
         year = 2025,
        month = feb,
       volume = {694},
          eid = {A184},
        pages = {A184},
          doi = {10.1051/0004-6361/202452477},
archivePrefix = {arXiv},
       eprint = {2410.02855},
 primaryClass = {astro-ph.GA},
       adsurl = {https://ui.adsabs.harvard.edu/abs/2025A&A...694A.184L}
}

@INPROCEEDINGS{Weinberg,
       author = {{Weinberg}, Martin D.},
        title = "{How Has Dynamical Evolution Changed the Galactic Globular Clusters?}",
    booktitle = {The Globular Cluster-Galaxy Connection},
         year = 1993,
       editor = {{Smith}, Graeme H. and {Brodie}, Jean P.},
       series = {Astronomical Society of the Pacific Conference Series},
       volume = {48},
        month = jan,
        pages = {689},
       adsurl = {https://ui.adsabs.harvard.edu/abs/1993ASPC...48..689W}
}

@ARTICLE{Keenan,
       author = {{Keenan}, D.~W. and {Innanen}, K.~A.},
        title = "{Numerical investigation of galactic tidal effects on spherical stellar systems.}",
      journal = {\aj},
         year = 1975,
        month = apr,
       volume = {80},
        pages = {290-302},
          doi = {10.1086/111744},
       adsurl = {https://ui.adsabs.harvard.edu/abs/1975AJ.....80..290K}
}

@ARTICLE{Ernst,
       author = {{Ernst}, A. and {Glaschke}, P. and {Fiestas}, J. and {Just}, A. and {Spurzem}, R.},
        title = "{N-body models of rotating globular clusters}",
      journal = {\mnras},
         year = 2007,
        month = may,
       volume = {377},
       number = {2},
        pages = {465-479},
          doi = {10.1111/j.1365-2966.2007.11602.x},
archivePrefix = {arXiv},
       eprint = {astro-ph/0702206},
 primaryClass = {astro-ph},
       adsurl = {https://ui.adsabs.harvard.edu/abs/2007MNRAS.377..465E}
}

@ARTICLE{2025A&A...697A...8M,
       author = {{Massari}, D. and {Dalessandro}, E. and {Erkal}, D. and {Balbinot}, E. and {Bovy}, J. and {McDonald}, I. and {Ferguson}, A.~M.~N. and {Larsen}, S.~S. and {Lan{\c{c}}on}, A. and {Annibali}, F. and {Goldman}, B. and {Kuzma}, P.~B. and {Voggel}, K. and {Saifollahi}, T. and {Cuillandre}, J.-C. and {Schirmer}, M. and {Kluge}, M. and {Altieri}, B. and {Amara}, A. and {Andreon}, S. and {Auricchio}, N. and {Baldi}, M. and {Balestra}, A. and {Bardelli}, S. and {Basset}, A. and {Bender}, R. and {Bonino}, D. and {Branchini}, E. and {Brescia}, M. and {Brinchmann}, J. and {Camera}, S. and {Candini}, G.~P. and {Capobianco}, V. and {Carbone}, C. and {Carlberg}, R.~G. and {Carretero}, J. and {Casas}, S. and {Castellano}, M. and {Cavuoti}, S. and {Cimatti}, A. and {Congedo}, G. and {Conselice}, C.~J. and {Conversi}, L. and {Copin}, Y. and {Corcione}, L. and {Courbin}, F. and {Courtois}, H.~M. and {Degaudenzi}, H. and {Dinis}, J. and {Dubath}, F. and {Dupac}, X. and {Dusini}, S. and {Fabricius}, M. and {Farina}, M. and {Farrens}, S. and {Ferriol}, S. and {Frailis}, M. and {Franceschi}, E. and {Garilli}, B. and {Gillis}, B. and {Giocoli}, C. and {Grazian}, A. and {Guzzo}, L. and {Hoar}, J. and {Hoekstra}, H. and {Holliman}, M.~S. and {Holmes}, W. and {Hook}, I. and {Hormuth}, F. and {Hornstrup}, A. and {Hudelot}, P. and {Jahnke}, K. and {Keih{\"a}nen}, E. and {Kermiche}, S. and {Kiessling}, A. and {Kitching}, T. and {Kohley}, R. and {Kubik}, B. and {K{\"u}mmel}, M. and {Kunz}, M. and {Kurki-Suonio}, H. and {Ligori}, S. and {Lilje}, P.~B. and {Lindholm}, V. and {Lloro}, I. and {Maino}, D. and {Maiorano}, E. and {Mansutti}, O. and {Marggraf}, O. and {Markovic}, K. and {Martinet}, N. and {Marulli}, F. and {Massey}, R. and {Maurogordato}, S. and {Medinaceli}, E. and {Mei}, S. and {Mellier}, Y. and {Meneghetti}, M. and {Meylan}, G. and {Moresco}, M. and {Moscardini}, L. and {Munari}, E. and {Nakajima}, R. and {Nichol}, R.~C. and {Niemi}, S.-M. and {Padilla}, C. and {Paltani}, S. and {Pasian}, F. and {Pedersen}, K. and {Percival}, W.~J. and {Pettorino}, V. and {Pires}, S. and {Polenta}, G. and {Poncet}, M. and {Popa}, L.~A. and {Pozzetti}, L. and {Racca}, G.~D. and {Raison}, F. and {Rebolo}, R. and {Renzi}, A. and {Rhodes}, J. and {Riccio}, G. and {Rix}, H.-W. and {Romelli}, E. and {Roncarelli}, M. and {Rossetti}, E. and {Saglia}, R. and {Sapone}, D. and {Sartoris}, B. and {Schneider}, P. and {Schrabback}, T. and {Secroun}, A. and {Seidel}, G. and {Seiffert}, M. and {Serrano}, S. and {Sirignano}, C. and {Sirri}, G. and {Skottfelt}, J. and {Stanco}, L. and {Tallada-Cresp{\'\i}}, P. and {Teplitz}, H.~I. and {Tereno}, I. and {Toledo-Moreo}, R. and {Torradeflot}, F. and {Tutusaus}, I. and {Valenziano}, L. and {Vassallo}, T. and {Veropalumbo}, A. and {Wang}, Y. and {Weller}, J. and {Zacchei}, A. and {Zamorani}, G. and {Zoubian}, J. and {Zucca}, E. and {Bolzonella}, M. and {Burigana}, C. and {Morris}, P.~W. and {Scottez}, V. and {Simon}, P. and {Mart{\'\i}n-Fleitas}, J. and {Scott}, D.},
        title = "{Euclid: Early Release Observations {\textendash} Unveiling the morphology of two Milky Way globular clusters out to their periphery}",
      journal = {\aap},
         year = 2025,
        month = may,
       volume = {697},
          eid = {A8},
        pages = {A8},
          doi = {10.1051/0004-6361/202449696},
archivePrefix = {arXiv},
       eprint = {2405.13498},
 primaryClass = {astro-ph.GA},
       adsurl = {https://ui.adsabs.harvard.edu/abs/2025A&A...697A...8M}
}



\appendix

\section{Details of asymptotic matching} \label{sec:A}
\subsection{Internal region} \label{sec:Ainternal}
First, we expand the internal solution $\psi^\text{(int)}$ and the dimensionless density $\hat{\rho}$ as perturbation series in tidal strength parameter $\epsilon$,
\begin{equation}
    \psi^\text{(int)}(\hat{\mathbfit{r}},t;\epsilon)=\sum_{\text{k=0}}^\infty\frac{1}{k!}\psi^\text{(int)}_k(\hat{\mathbfit{r}},t)\epsilon^k,
\end{equation}
\begin{equation}
    \hat{\rho}(\hat{\mathbfit{r}},t;\epsilon)=\sum_{k=0}^\infty\frac{1}{k!}\hat{\rho}_k(\hat{\mathbfit{r}},t)\epsilon^k.
\end{equation}
Inserting these expansions into the dimensionless Poisson equation~(\ref{eq:poisson}), for $k=0$, we obtain the ordinary differential equation that governs the spherically symmetric \cite{King} models (equation~(\ref{eq:kingpoisson})). The central boundary conditions (\ref{eq:bc1})-(\ref{eq:bc2}) are applied to give $\psi_0^{\text{(int)}}(0)=\Psi$ and $\psi_0^{\text{(int)}\prime}(0)=0$. The solution can be obtained numerically, using standard methods (see Appendix~\ref{sec:BTIRO} for further details about the numerical implementation that accompanies this article).

For $k=1$, we obtain the partial differential equation
\begin{equation}
    \left[\hat{\nabla}^2+R_1(\hat{r};\Psi)\right]\psi_1^{\text{(int)}}(\hat{\mathbfit{r}},t)=-9(1-\nu-2\zeta), \label{eq:k1}
\end{equation}
where the quantity $R_1$ is defined as
\begin{equation}
    R_1(\hat{r};\Psi)=\frac{9}{\hat{\rho}(\Psi)}\frac{d\hat{\rho}}{d\psi}\bigg\rvert_{\psi^\text{(int)}_0} = \frac{9}{\hat{\rho}(\Psi)}\frac{\hat{\rho}_1}{\psi_1^\text{(int)}}.
\end{equation}
 Since we chose to apply $\psi_0^\text{(int)}(0)=\Psi$, the boundary conditions at higher orders are homogeneous: $\psi_1^{\text{(int)}}(\mathbfit{0},t)=0$ and $\hat{\nabla}\psi_1^{\text{(int)}}(\mathbfit{0},t)=\mathbfit{0}$. To separate the radial and angular components of the $k=1$ solution, we expand in spherical harmonics,
\begin{equation}
    \psi_1^\text{(int)}(\hat{\mathbfit{r}},t)=\sum_{l=0}^\infty\sum_{m=-1}^l\psi_{1,lm}^\text{(int)}(\hat{r})Y_{lm}(\theta,\phi).
\end{equation}
Using the standard result
\begin{equation}
    \hat{\nabla}^2Y_{lm}(\theta,\phi)= -\frac{l(l+1)}{\hat{r}^2}Y_{lm}(\theta,\phi),
\end{equation}
we may define a linear differential operator $\mathcal{D}_l$,
\begin{equation}
    \mathcal{D}_l=\frac{d^2}{d\hat{r}^2}+\frac{2}{\hat{r}}\frac{d}{d\hat{r}}-\frac{l(l+1)}{\hat{r}^2}+R_1(\hat{r};\Psi), \label{eq:Dl}
\end{equation}
with which we can rewrite equation~(\ref{eq:k1}) as
\begin{equation}
    \sum_{l=0}^\infty\sum_{m=-1}^l Y_{lm}(\theta,\phi)\mathcal{D}_l\psi^\text{(int)}_{1,lm}(\hat{r}) = -9(1-\nu-2\zeta).
\end{equation}
Choosing to apply the constant terms at $l=0$, we obtain the set of ordinary differential equations~(\ref{eq:ode1})--(\ref{eq:ode2}). The boundary conditions for all harmonics are homogeneous.

In the limit $\hat{r}\!\rightarrow\!0$, the differential equation governing harmonics of degree $l\geq1$ reduces to Bessel's equation \citep[see][]{NIST:DLMF}. In this limit, the Bessel's equation has two independent solutions, which behave as $\hat{r}^l$ and $\hat{r}^{-(l+1)}$. Since only $\hat{r}^l$ satisfies $\psi_{1,lm}^\text{(int)}(0)=0$, we can deduce that, for $l\geq1$,
\begin{equation}
    \psi_{1,lm}^\text{(int)}(\hat{r})=A_{lm}\mathcal{\gamma}_l(\hat{r}),
\end{equation}
where $\mathcal{\gamma}_l(\hat{r})\sim\hat{r}^l$ as $\hat{r}\!\rightarrow\!0$ and $A_{lm}$ are unknown constants. By the same argument, there can be no homogeneous solution for $l=0$ as $\hat{r}^0=1$ cannot satisfy $\psi^\text{(int)}_{1,00}(0)=0$.

\subsection{Boundary layer} \label{sec:Aboundary}
The validity of the perturbation expansion in equation~(\ref{eq:expansion}) breaks down when the unperturbed term becomes comparable to the linear term: $\psi_0\!=\!\mathcal{O}(\epsilon)$. Since the first non-vanishing term in the Taylor expansion of $\psi_0$ about the truncation radius $r_\text{tr}$ is linear in $(\hat{r}-\hat{r}_\text{tr})$, we define the boundary layer as the region in which $\hat{r}_\text{tr}-\hat{r}=\mathcal{O}(\epsilon)$. Rescaling $\psi^\text{(lay)}$ as in equation~(\ref{eq:tau}), the Poisson equation~(\ref{eq:newpoisson}) becomes \begin{equation}
    \epsilon\hat{\nabla}^2\tau= -9\left[\frac{\hat{\rho}(\epsilon\tau)}{\hat{\rho}(\Psi)}+(1-\nu-2\zeta)\epsilon\right]. \label{eq:bpoisson}
\end{equation}
The asymptotic behaviour of $\hat{\rho}(\psi)$ (see equation~(\ref{eq:density})) for small arguments is
\begin{equation}
    \hat{\rho}(\epsilon\tau)
    \sim \frac{2}{5}(\epsilon\tau)^{5/2}+\frac{4}{35}(\epsilon\tau)^{7/2}+\mathcal{O}(\epsilon^{9/2}).
\end{equation}
Consequently, after rescaling the radial coordinate as in equation~(\ref{eq:eta}), the boundary layer Poisson equation~(\ref{eq:bpoisson}) reduces to the Laplace equation given in equation~(\ref{eq:pl}), up to third order in $\epsilon$. Since the right-hand side of this equation is second order in $\epsilon$, the differential equations for zeroth and first order terms are simply
\begin{equation}
    \frac{\partial^2\tau_0}{\partial\eta^2}=0,
\end{equation}
\begin{equation}
    \frac{\partial^2\tau_1}{\partial\eta^2}=\frac{2}{\hat{r}_\text{tr}}\frac{\partial\tau_0}{\partial\eta}.
\end{equation}
The former may be directly integrated to give equation~(\ref{eq:tau0}), while the latter is solved by a function of the form $\tau_1=\tau_0\eta\hat{r}_\text{tr}^{-1}+D(\theta,\phi)\eta+E(
\theta,\phi)$. Substituting in for $\tau_0$ and relabelling angular functions gives equation~(\ref{eq:tau1}).

\subsection{Asymptotic Matching} \label{sec:Amatching}
Asymptotic matching between the external and boundary layer solutions is conducted according to the \cite{VanDyke} principle (see equation (5.24) therein). First, the external solution must be expanded in a Taylor series about $\hat{r}=\hat{r}_\text{tr}$, which for $k=0$ gives
\begin{equation}
    \psi_0^\text{(ext)}(\hat{r})=\alpha_0-\frac{\lambda_0}{\hat{r}_\text{tr}}+\frac{\lambda_0}{\hat{r}_\text{tr}^2}(\hat{r}-\hat{r}_\text{tr}),
\end{equation}
up to order $\mathcal{O}(\hat{r}-\hat{r}_\text{tr})$. By comparison with equation~(\ref{eq:layer}), the vanishing $\psi_0^\text{(lay)}(\hat{r}_\text{tr})$ term leads to the definition of $\alpha_0$ in equation~(\ref{eq:alpha0}). Similarly, we can identify $F_0(\theta,\phi)=-\lambda_0/\hat{r}_\text{tr}^2$, which couples with equation~(\ref{eq:F0}) to determine $\lambda_0$ as in equation~(\ref{eq:lambda0}).

Using the form of the $k=1$ internal solution given in equation~(\ref{eq:internal}), the boundary layer angular functions given in equations~(\ref{eq:G0})--(\ref{eq:F1}) can be expressed as
\begin{equation}
G_0(\theta,\phi)=f_{00}(\hat{r}_\text{tr})+\sum_{l=1}^\infty\sum_{m=-l}^l A_{lm}\mathcal{\gamma}_l(\hat{r}_\text{tr})Y_{lm}(\theta,\phi), \label{eq:matchA}
\end{equation}
\begin{equation}
F_1(\theta,\phi)=-f_{00}'(\hat{r}_\text{tr})-\sum_{l=1}^\infty\sum_{m=-l}^l A_{lm}\mathcal{\gamma}_l'(\hat{r}_\text{tr})Y_{lm}(\theta,\phi). \label{eq:matchB}
\end{equation}
By matching with the $k=1$ external solution (see equation~(\ref{eq:external})), Taylor expanded about $\hat{r}=\hat{r}_\text{tr}$, we obtain
\begin{align}
    G_0(\theta,&
    \phi)= \alpha_1-\frac{\lambda_1}{\hat{r}_\text{tr}}-\frac{T_{00}(\hat{r}_\text{tr})+\zeta C_{00}(\hat{r}_\text{tr})}{\sqrt{4\pi}} \notag\\&-\sum_{l=1}^{\infty}\sum_{m=-l}^l\left[\frac{a_{lm}}{\hat{r}_\text{tr}^{l+1}}+T_{lm}(\hat{r}_\text{tr})+\zeta C_{lm}(\hat{r}_\text{tr})\right]Y_{lm}(\theta,\phi), \label{eq:matcha}
\end{align}
\begin{align}
    F_1(\theta,&\phi)=-\frac{\lambda_1}{\hat{r}_\text{tr}^2} +\frac{T_{00}'(\hat{r}_\text{tr})+\zeta C_{00}'(\hat{r}_\text{tr})}{\sqrt{4\pi}}\notag\\&+\sum_{l=1}^{\infty}\sum_{m=-l}^l\left[-(l+1)\frac{a_{lm}}{\hat{r}_\text{tr}^{l+2}}+T_{lm}'(\hat{r}_{tr})+\zeta C_{lm}'(\hat{r}_{tr})\right]Y_{lm}(\theta,\phi). \label{eq:matchb}
\end{align}
Equating coefficients between equations~(\ref{eq:matchA}) and (\ref{eq:matcha}) and between equations~(\ref{eq:matchB}) and (\ref{eq:matchb}), we find
\begin{equation}
    \alpha_1-\frac{\lambda_1}{\hat{r}_\text{tr}}-\frac{T_{00}(\hat{r}_\text{tr})+\zeta C_{00}(\hat{r}_\text{tr})}{\sqrt{4\pi}}=f_{00}(\hat{r}_\text{tr}) \label{eq:motch}
\end{equation}
\begin{equation}
    \frac{a_{lm}}{\hat{r}_\text{tr}^{l+1}}+T_{lm}(\hat{r}_\text{tr})+\zeta C_{lm}(\hat{r}_\text{tr})=-A_{lm}\mathcal{\gamma}_l(\hat{r}_\text{tr}), \label{eq:metch}
\end{equation}
\begin{equation}
    \frac{\lambda_1}{\hat{r}_\text{tr}^2} -\frac{T_{00}'(\hat{r}_\text{tr})+\zeta C_{00}'(\hat{r}_\text{tr})}{\sqrt{4\pi}}=f_{00}'(\hat{r}_\text{tr}), \label{eq:mitch}
\end{equation}
\begin{equation}
    (l+1)\frac{a_{lm}}{\hat{r}_\text{tr}^{l+2}}-T_{lm}'(\hat{r}_{tr})-\zeta C_{lm}'(\hat{r}_{tr})=A_{lm}\mathcal{\gamma}_l'(\hat{r}_\text{tr}), \label{eq:mutch}
\end{equation}
where $l\geq1$. Equation~(\ref{eq:mitch}) rearranges to give equation~(\ref{eq:lambda1}) for $\lambda_1$, which is inserted into equation~(\ref{eq:motch}) to obtain equation~(\ref{eq:alpha1}) for $\alpha_1$. Equations~(\ref{eq:metch}) and (\ref{eq:mutch}) combine to give
\begin{equation}
    A_{lm}=-\frac{(l+1)\big[T_{lm}(\hat{r}_\text{tr})+\zeta C_{lm}(\hat{r}_\text{tr})\big]+\hat{r}_\text{tr}\big[T_{lm}'(\hat{r}_\text{tr})+\zeta C'_{lm}(\hat{r}_\text{tr})\big]}{\hat{r}_\text{tr}\mathcal{\gamma}_l'(\hat{r}_\text{tr})+(l+1)\mathcal{\gamma}_l(\hat{r}_\text{tr})},
\end{equation}
which is nonvanishing for $l=2$ with $m\in\{0,2\}$ only. By using equations~(\ref{eq:T00})--(\ref{eq:C20}) to find $T'_{lm}(\hat{r})$ and $C'_{lm}(\hat{r})$, we determine constants $a_{20}$, $a_{22}$, $A_{20}$ and $A_{22}$ as in equations (\ref{eq:a20})-(\ref{eq:A22}). $A_{lm}\!=\!a_{lm}\!=\!0$ in all other cases. 

\section{The TIRO code}
\label{sec:BTIRO}
The Poisson solver used in the construction of these models was written using Python 3.13.5. The code is called TIdal ROtational solver (TIRO). The required inputs are $\Psi$, $\epsilon$, $\zeta$ and $\nu$. Numerical integration of equations (\ref{eq:kingpoisson}), (\ref{eq:ode1}) and (\ref{eq:ode2}) is performed using an explicit Runge-Kutta method of order 8 ('DOP853'), implemented using the \verb|solve_ivp| function from the \verb|scipy.integrate| module. The lower integration limit is taken to be at a small offset from $\hat{r}=0$, which is corrected for using the Frobenius method. The integration results are combined with computed values of relevant constants to obtain the internal and external solutions for $\psi$. These are joined at the truncation radius to obtain the full global solution. The tidal radius is found using the \verb|brentq| root finder from the \verb|scipy.optimize| module. Normalised density and velocity dispersion profiles are outputted, alongside slices through the equipotential surfaces and values for the tidal and truncation radii. Modules are provided to convert outputs from dimensionless coordinates into N-body or physical units. The TIRO code is available at: \url{https://github.com/lucyarditi/TIRO.git}.


\bsp	
\label{lastpage}
\end{document}